\documentclass[aps,pra,10pt,twocolumn,notitlepage,superscriptaddress,dvipsnames,longbibliography, tightenlines,floatfix]{revtex4-1}
\pdfoutput=1
\usepackage{xcolor}
\usepackage{bm}
\usepackage{bbold}
\usepackage{mathtools}
\usepackage{graphicx}
\usepackage[caption=false, subrefformat=parens, labelformat=parens]{subfig}
\usepackage{enumitem}
\setlist{leftmargin=.8cm}
\usepackage{float}
\usepackage{textcomp}
\usepackage{xcolor}
\definecolor{dark-red}{rgb}{0.4,0.15,0.15}
\definecolor{dark-blue}{rgb}{0.15,0.15,0.4}
\definecolor{medium-blue}{rgb}{0,0,0.5}
\usepackage{hyperref}
\usepackage[all]{hypcap}%fix problems with figures and tables of the 'hyperref' package
\usepackage{cleveref}% Package hperref works wrongly within the 'subequations' environment without this.
\hypersetup{
    colorlinks, linkcolor={dark-red},
    citecolor={dark-blue}, urlcolor={medium-blue}
} %These setups replace the ugly and distracting red and green boxes of the 'hyperref' package with colored texts.
\newcommand{\nsp}{\hspace{-0.4pt}}
\newcommand{\ssp}{\hspace{0.4pt}}
\newcommand{\norm}[1]{\lvert #1 \rvert}

\newcommand{\ket}[1]{\lvert\, #1\, \rangle}
\newcommand{\ketb}[1]{\big\lvert\, #1\, \big\rangle}
\newcommand{\bra}[1]{\langle\, #1\, \rvert}

\newcommand{\commut}[2]{{[\, #1,\, #2 \,]}}

\newcommand{\anticommut}[2]{{\{\, #1,\, #2 \,\}}}
\newcommand{\half}{\frac{1}{2}}
\newcommand{\sqhalf}{\frac{1}{\sqrt 2}}
\newcommand{\hc}{\mathrm{H.c.}}
\newcommand{\vac}{\mathrm{vac}}
\newcommand{\anc}{a}
\newcommand{\sys}{s}
\newcommand{\bnc}{b}
\newcommand{\tys}{t}
\newcommand{\nn}{n}
\newcommand{\num}{n}

\newcommand{\ea}{{\it et al.}}
\newcommand{\cnumber}{\text{$c$ number}}
\newcommand{\azero}{\bm}
\newcommand{\azerotext}{bold}
\newcommand{\pzero}{\underline}
\newcommand{\pzerotext}{underlined}

\DeclareMathOperator{\diag}{diag}
\DeclareMathOperator{\sgn}{sgn}

\newcommand{\xx}{*}

\renewcommand{\c}{c}
\renewcommand{\b}{b}

\newcommand{\mm}{M}
\newcommand{\bF}[1]{\mathcal F^b_{\raisebox{1.4pt}{$\scriptstyle #1$}}}
\newcommand{\bK}{K}
\newcommand{\s}{f}
\newcommand{\SWAP}{{\footnotesize SWAP}}
\newcommand{\python}{{\footnotesize PYTHON}}
\newcommand{\FSWAP}{{\footnotesize FSWAP}}
\newcommand{\ISWAP}{{\footnotesize ISWAP}}
\newcommand{\mSWAP}{{\mathrm{SWAP}}}
\newcommand{\mISWAP}{{\mathrm{ISWAP}}}
\newcommand{\tran}{\mathcal U}
\newcommand{\orth}{R}
\newcommand{\ph}{B}

\newcommand{\nso}{N} % number of spin orbitals
\newcommand{\nel}{N_{\nsp f}} % number of electrons
\newcommand{\kbf}{{\mathbf k}}
\newcommand{\sysbf}{{\mathbf s}}
\newcommand{\CZ}{{\footnotesize{\rm CZ}}}
\newcommand{\CNOT}{{\footnotesize{\rm CNOT}}}
\mathchardef\myhyphen="2D % Define a "math hyphen"

\newcommand{\Hhop}{\mathcal H_{\text{hop}}}
\newcommand{\Hint}{\mathcal H_{\text{int}}}

\begin{document}

\title{Quantum Algorithms to Simulate Many-Body Physics of Correlated Fermions}
\date{\today}
\author{Zhang Jiang}
\email[Present address: Google, Venice, California 90291, USA.\\
Corresponding author.\\]{
qzj@google.com}
\affiliation{NASA Ames Research Center Quantum Artificial Intelligence Laboratory (QuAIL),  Moffett Field, CA 94035, USA}
\affiliation{Stinger Ghaffarian Technologies Inc., 7701 Greenbelt Road, Suite~400, Greenbelt, MD 20770, USA}

\author{Kevin J. Sung}
\affiliation{Department of Electrical Engineering and Computer Science, University of Michigan, Ann Arbor, MI 48109, USA}

\author{Kostyantyn Kechedzhi}
\affiliation{NASA Ames Research Center Quantum Artificial Intelligence Laboratory (QuAIL),  Moffett Field, CA 94035, USA}
\affiliation{USRA, NASA Ames Research Center, Moffett Field, CA 94035}

\author{Vadim N. Smelyanskiy}
\affiliation{Google, Venice, CA 90291, USA}

\author{Sergio Boixo}
\affiliation{Google, Venice, CA 90291, USA}

\begin{abstract}

Simulating strongly correlated fermionic systems is notoriously hard on classical computers. An alternative approach, as proposed by Feynman, is to use a quantum computer. We discuss simulating strongly correlated fermionic systems using near-term quantum devices. We focus specifically on two-dimensional (2D) or linear geometry with nearest-neighbor qubit-qubit couplings, typical for superconducting transmon qubit arrays. We improve an existing algorithm to prepare an arbitrary Slater determinant by exploiting a unitary symmetry. We also present a quantum algorithm to prepare an arbitrary fermionic Gaussian state with $O(\nso^2)$ gates and $O(\nso)$ circuit depth. Both algorithms are optimal in the sense that the numbers of parameters in the quantum circuits are equal to those describing the quantum states. Furthermore, we propose an algorithm to implement the 2D fermionic Fourier transformation on a 2D qubit array with only $O(\nso^{1.5})$ gates and $O(\sqrt \nso)$ circuit depth, which is the minimum depth required for quantum information to travel across the qubit array. We also present methods to simulate each time step in the evolution of the 2D Fermi-Hubbard model---again on a 2D qubit array---with $O(\nso)$ gates and $O(\sqrt \nso)$ circuit depth. Finally, we discuss how these algorithms can be used to determine the ground-state properties and phase diagrams of strongly correlated quantum systems using the Hubbard model as an example.

\end{abstract}

\maketitle

\section{Introduction}
\label{sec:intro}

A large class of materials evade description by density functional theory~\cite{hohenberg_inhomogeneous_1964} due to the effects of strong electron-electron correlations~\cite{quintanilla_strong-correlations_2009,basov_electrodynamics_2011}. A simulation of strongly correlated electronic structure and dynamical effects in such materials would allow a quantitative prediction of their physical characteristics before fabricating them, which is often costly. This approach would open the route for designing materials with application-specific characteristics. The properties of strongly correlated fermionic systems remain elusive after many years of intensive research. Indeed, solving a general quantum many-body problem without using approximations on a classical computer takes exponential time in the size of the problem. 

One way to avoid this difficulty, as envisioned by Feynman, is to use a quantum computer to simulate quantum systems~\cite{feynman_simulating_1982,georgescu_quantum_2014}. In experiments with real materials, one rarely knows for sure the initial states and the underlying Hamiltonians. A quantum computer, however, enables one to prepare the initial state with confidence and to have full control over the Hamiltonian under which the state evolves. We are closer to Feynman's vision with recent advances in quantum-computing hardware. It is important to develop simulation algorithms that are optimized given the limitations of current and near-term quantum hardware, such as the locality of qubit-qubit couplings, the gate sequences, and the mitigation of errors and noise. Significant progress has been made in this direction~\cite{hastings_improving_2015,kandala_hardware-efficient_2017,reiher_elucidating_2017,babbush_low_2017}. 

We propose several quantum algorithms to simulate correlated fermions on 2D and linear qubit arrays with nearest-neighbor couplings which are typical for superconducting transmon qubits. We use the Fermi-Hubbard model as an example to demonstrate these algorithms. In this simple model, the superconducting-to-Mott-insulator transition is solely determined by the competition between the hopping and the interaction terms. Solving the Hubbard model allows one to single out this mechanism from other effects such as disorder and long-range interactions in real materials.

The Hubbard model~\cite{hubbard_electron_1963} approximates the long-range Coulomb interaction of electrons in a crystal with a local on-site interaction. This locality reduces the resources required for simulating the model and makes it a prime candidate for the early applications of quantum simulations~\cite{wecker_solving_2015}. The single-band Fermi-Hubbard model is described by the Hamiltonian
\begin{align}\label{eq:fhh}
\mathcal H_\mathrm{FH} = &- \sum_{\langle j,k \rangle, \sigma} t_{jk}\big( \c^\dagger_{j,\sigma} \c^{}_{k,\sigma} +\hc\big) +U \sum_j \num_{j,\uparrow}\num_{j,\downarrow}\nonumber\\
&\;\;\; + \sum_{j,\sigma} (\epsilon_j-\mu) \num_{j,\sigma}- \sum_j h_j (\num_{j,\uparrow}-\num_{j,\downarrow})\,,
\end{align}
where $\c^\dagger_{j,\sigma}$ ($\c_{j,\sigma}$) is the creation (annihilation) operator for the $j$th site with spin $\sigma$ and
$\num_{j,\sigma} = \c^\dagger_{j,\sigma} \c^{}_{j,\sigma}$ is the fermion occupation number operator. The first term on the right side of Eq.~\eqref{eq:fhh} describes fermions hopping between sites, the second term describes the on-site interactions, and the remaining two terms describe a local potential and a magnetic field. The model demonstrates a wide range of strongly correlated phenomena, including metal-insulator transitions, unconventional Fermi liquids, and a number of inhomogeneous phases. The Fermi-Hubbard model also provides an approximate description of materials~\cite{1bFHMmaterial} including the cuprate family~\cite{ShluterCuprateParameters} (albeit a multiband extension of the model is necessary for quantitative correspondence to the materials), which has attracted a lot of interest because of unconventional symmetry-breaking phenomena and high-temperature superconductivity~\cite{dagotto_correlated_1994}.

The one-dimensional (1D) Hubbard model was solved by Lieb and Wu in 1968~\cite{Lieb_absence_1968}; however, a full theoretical analysis of the 2D Hubbard model requires going beyond the validity of mean-field and perturbation theory arguments remains an open question. The 2D Hubbard model serves as a canonical microscopic physical model for strongly correlated fermionic systems. In the underdoped region of its phase diagram, multiple orders exist corresponding to a regime of maximum numerical difficulty. Significant numerical progress has been made in the identification of the unconventional Mott-insulator transition and a superconducting phase with a $d$-wave order parameter~\cite{maier_d-wave_2000,zheng_ground-state_2016}. Exact numerical diagonalization is limited to about 20 sites (40 logical qubits)~\cite{PRB2011FidelityNumerics,Trapped2DFHnumericsED}, which is too small for a finite-size scaling analysis. Approximate methods such as quantum Monte Carlo simulations or many-body theory expansions have been used to simulate systems with hundreds of sites, which allows for extrapolation to the thermodynamic limit. Except at half filling, quantum Monte Carlo methods suffer from sign problems which prevent them from simulating systems at very low temperatures~\cite{troyer_computational_2005}. Application of the density matrix renormalization group (DMRG)~\cite{white_density-matrix_1993} to the 2D Hubbard model requires mapping to an effective 1D problem, to which the standard DMRG is applied. Other numerical methods to determine the phase diagrams of the Hubbard model include the dynamical cluster approximation~\cite{hettler_dynamical_2000,kotliar_electronic_2006} and the density matrix embedding theory~\cite{knizia_density_2012,zheng_ground-state_2016}. These methods can asymptotically approach the exact solution with increasingly larger clusters; however, simulating those clusters requires an exponential amount of computing resources on a classical computer. 

To simulate the Fermi-Hubbard model on a quantum computer, one needs to map the fermionic operators to qubit operators. In the second quantization picture, a particular spin orbital being unoccupied (occupied) can be represented by the qubit state $\ket{0}$ ($\ket{1}$). To retain the fermionic anticommutation relations, one also needs to account for the parities of qubits corresponding to other spin orbitals, e.g., by using the Jordan-Wigner transformation (JWT)~\cite{ortiz_quantum_2001,whitfield_simulation_2011}. The two terms on the second line of Eq.~\eqref{eq:fhh} can be implemented using only single-qubit operators, and the on-site interaction term can be implemented with two-qubit interactions. The hopping terms, however, cannot be implemented straightforwardly in more than one spatial dimension because of the nonlocal parity operators in the JWT. It is of practical importance to reduce the depth of the quantum circuits for these terms with only local qubit interactions~\cite{babbush_low_2017}, which is crucial to near-term quantum computers without quantum error correction~\cite{houck_on-chip_2012,barends_digitized_2016}. We will always have local qubit interactions in mind, which is a different model from quantum computation with all-to-all interactions. The digital quantum simulations we consider here could address low-temperature properties of the Hubbard model, which thus far remain hard to access with analog simulators~\cite{mazurenko_cold-atom_2017}. In addition to simulating unitary dynamics, preparing the ground state is a source of significant overhead in digital quantum simulations. We develop algorithms for state preparations with only locally coupled qubits by adopting the general ideas from Ref.~\cite{wecker_solving_2015}.

Beyond the Hubbard model, there are a growing number of materials showing surprising sensitivity to impurities. It is an open issue to understand and, ultimately, to control the emergent phases from impurities~\cite{seo_disorder_2013}. At low temperatures, impurities and disorder can induce a metal-to-insulator transition by localizing the Cooper pairs~\cite{sacepe_localization_2011,kondov_disorder-induced_2015}. Impurities, however, can also induce interesting phases of strongly correlated matter and help probe the underlying mechanism of exotic states~\cite{zhu_spin_2002,andersen_disorder-induced_2007}. 
% Materials that exhibit an extreme sensitivity to impurities often are near a zero-temperature, second-order phase transition, where quantum critical fluctuations of the associated order parameter diverge in space and time. Theory predicts that when disorder is coupled to these critical fluctuations local regions of an exotic phase can nucleate inside the host and change the nature of the phase transition; indeed, even long-range order of the nucleated phase is possible~\cite{zhu_spin_2002,andersen_disorder-induced_2007}. Nuclear magnetic resonance and scanning tunnelling microscopy have validated these predictions by showing that the host system locally responds to an impurity by creating an extended droplet of the new phase around an impurity. Understanding how the local droplet evolves as a function of a control parameter or how the droplet interplays with the host phase, however, is less well known, partly owing to a lack of very pure compounds that are situated sufficiently close to a QCP.
On the other hand, quantum impurity models are useful for hybrid quantum-classical approaches to strongly correlated materials~\cite{bauer_hybrid_2016}, where the quantum computer solves an impurity problem that is determined self-consistently with the help of a classical computer. Our methods work for systems that violate translational symmetries and thus are suited for simulating impurity models. 

In this paper, we propose several quantum algorithms to simulate fermionic systems on near-term devices, e.g., on 2D or linear geometry with nearest-neighbor qubit-qubit couplings. Specifically, we propose quantum algorithms to prepare arbitrary fermionic Gaussian states which can be used as a starting point to model correlated quantum states. The same algorithms can also be used to implement arbitrary fermionic transformations with linear input-output relations. We also provide a state-of-the-art algorithm for 2D fermionic Fourier transformation, where the fermionic parity problem is overcome with negligible overhead. Our algorithms open up the possibility of a wealth of experiments to simulate many-body physics, including the Fermi-Hubbard model, in the noisy
intermediate-scale quantum era~\cite{preskill_quantum_2018}.

The paper is organized in the following way. In Sec.~\ref{sec:mapping}, we review the topic of mapping fermionic operators to qubit operators and introduce notations. In Sec.~\ref{sec:slater}, we improve an existing quantum algorithm in Ref.~\cite{wecker_solving_2015} to prepare an arbitrary Slater determinant by exploiting a unitary symmetry. In Sec.~\ref{sec:gaussian}, we present an algorithm to prepare an arbitrary fermionic Gaussian state, improving on and generalizing an existing method for translationally invariant systems~\cite{verstraete_quantum_2009}. Both algorithms in Secs.~\ref{sec:slater} and \ref{sec:gaussian} are implemented as part of the open-source project OpenFermion~\cite{mcclean_openfermion_2017}. In Sec.~\ref{sec:fourier}, we introduce a quantum algorithm to implement the 2D fermionic Fourier transformation on a 2D qubit array with only $O(\nso^{1.5})$ gates and $O(\sqrt \nso)$ circuit depth, a better scaling than methods based on the fermionic \SWAP\ gates~\cite{babbush_low_2017}. In Sec.~\ref{sec:fhm}, we discuss applications of our quantum algorithms using the Fermi-Hubbard model as an example. In Appendix~\ref{sec:quadratic}, we review the properties of mean-field Hamiltonians that are quadratic in the fermionic creation and annihilation operators and discuss how to diagonalize these Hamiltonians. In Appendix~\ref{sec:move_ancilla}, we introduce an alternative approach to solve time evolutions of the quantum lattice models, such as the Hubbard model, by using ancilla-assisted fermionic gates. In Appendix~\ref{sec:hamiltonian_based}, we discuss fermionic operations based on Hamiltonian evolution. In Appendix~\ref{sec:FSWAP}, we briefly review the fermionic \SWAP\ gate. In Appendix~\ref{sec:qubit_ladder}, we discuss how to simulate the 2D Hubbard model with a qubit ladder using the fermionic \SWAP\ gate. In Appendix~\ref{sec:n_trotter}, we study the Trotter errors in adiabatic state preparation numerically for small system sizes.

\section{Mapping fermions to qubits}
\label{sec:mapping}

The first step in solving strongly correlated fermionic systems with a quantum computer is to map the Hilbert space of fermions to the states of qubits. Following Ref.~\cite{lanyon_towards_2010}, we represent the fermionic Hamiltonian in the second quantization picture using a discrete basis of spin orbitals. A qubit is assigned to each spin orbital; the qubit state $\ket{0}$ ($\ket{1}$) denotes an unoccupied (occupied) spin orbital. The fermionic operators satisfying the anticommutation relations can be mapped to qubit operators using the JWT~\cite{ortiz_quantum_2001,whitfield_simulation_2011}, the
Bravyi-Kitaev transformation (BKT)~\cite{bravyi_fermionic_2002,seeley_bravyi-kitaev_2012}, or the Ball-Verstraete-Cirac transformation (BVCT)~\cite{ball_fermions_2005,verstraete_mapping_2005,whitfield_local_2016}. To represent the parities, the JWT requires strings of Pauli operators that act on $O(\nso)$ qubits, the BKT uses $O(\log\nso)$ nonlocal operators, and the BVCT requires only $O(1)$ local operators by introducing one ancilla qubit per logical qubit. Although the JWT seems to have the worst scaling, it has the most straightforward form and can be applied to 1D fermionic systems without encountering the parity problem. One way to get around the parity problem in the JWT for systems of higher dimensions is to use fermionic \SWAP\ gates~\cite{babbush_low_2017,kivlichan_quantum_2017}. We discuss in this paper a different approach based on traveling ancilla qubits, 
which solves the parity problem in the JWT with negligible overhead.

The JWT maps the fermionic creation and annihilation operators to qubit operators as follows:
\begin{subequations}\label{eq:jw}
\begin{gather}
 \c^\dagger_{j} \;\mapsto\; \half\, \big(X_j-iY_j\big)\, Z_1\cdots Z_{j-1}\,,\\[3pt]
 \c_{j} \;\mapsto\; \half\, \big(X_j+iY_j\big)\, Z_1\cdots Z_{j-1}\,,
\end{gather}
\end{subequations}
where $X$, $Y$, and $Z$ are the Pauli operators. It assumes an ordering of the qubits and attaches the Pauli-$Z$ operators with smaller indices (the parity) to the raising and lowering operators. A single Slater determinant in the computational basis (Fock state) takes the form
\begin{align}\label{eq:hatree_fock}
 \c^\dagger_{j_1}\cdots\c^\dagger_{j_{\nel}}\ket{\vac} \;\mapsto\; \left(\prod_{n=1}^{\nel} \frac{X_{j_n}-iY_{j_n}}{2}\right) \ketb{0\cdots 0}\,,
\end{align}
where $j_1  < j_2\cdots < j_{\nel}$ and $\ket{\vac}$ is the vacuum state. The fermionic number operator
\begin{align}
 \c^\dagger_{j}\ssp\c_j \;\mapsto\; \frac{1}{4}\, \big(X_j-iY_j\big)\big(X_j+iY_j\big)=\half \,\big(I-Z_j\big)\,,
\end{align}
where $I$ is the $2\times 2$ identify operator. The fermionic hopping term can be realized by a product of Pauli operators ($k>j$),
\begin{align}\label{eq:hopping}
 \c^\dagger_{j}\ssp\c_k \;\mapsto\; \frac{1}{4}\, \big(X_j-iY_j\big)\big(X_k+iY_k\big)\, Z_{j+1}\cdots Z_{k-1}\,.
\end{align}
For the case $k=j+1$, we have
\begin{subequations}\label{eq:n_hopping}
\begin{gather}
-i\ssp\big(\c^\dagger_{j}\ssp\c_{j+1} - \hc\big)  
 \;\mapsto\; \frac{1}{2}\ssp \big(X_j Y_{j+1} - Y_j X_{j+1}\big)\label{eq:n_hopping_a}
\,,\\[3pt]
\label{eq:n_hopping_b}
 \c^\dagger_{j}\ssp\c_{j+1} + \hc 
 %& \frac{1}{4}\, \big(X_j-iY_j\big)\big(X_{j+1}+iY_{j+1}\big) +\hc\\
 \;\mapsto\; \frac{1}{2}\ssp \big(X_j X_{j+1} + Y_j Y_{j+1}\big)\,, 
\end{gather}
\end{subequations}
interactions which can be implemented efficiently with superconducting qubits~\cite{wendin_quantum_2017}. The hopping terms between orbitals encoded far from each other in the Jordan-Wigner transformation, however, are generally hard to implement. In Secs.~\ref{sec:slater} and \ref{sec:gaussian}, we discuss state preparation algorithms that totally avoid such hopping terms. In Sec.~\ref{sec:fourier} and Appendix~\ref{sec:move_ancilla}, we also discuss strategies to implement such terms by introducing ancilla qubits that store the parities.

\section{Preparing Slater determinants}
\label{sec:slater}

A Slater determinant can be regarded as an eigenstate of a Hamiltonian quadratic in fermionic creation and annihilation operators, or simply as a quadratic Hamiltonian. The standard algorithm to prepare Slater determinants was described in Ref.~\cite{ortiz_quantum_2001} and improved in Ref.~\cite{wecker_solving_2015} using elementary operations called Givens rotations, which are rotations in the plane spanned by two coordinates axes. The circuit depth and implementability can be improved by parallelizing the Givens rotations restricted to neighboring qubits~\cite{kivlichan_quantum_2017}. Here, we present an algorithm that reduces the total number of Givens rotations by exploiting a freedom in the representation of Slater determinants; \python\ code for this algorithm is available as part of the OpenFermion project~\cite{mcclean_openfermion_2017}.

In the second quantization picture, a single Slater determinant takes the form
\begin{align}\label{eq:slater}
 \ket{\Psi_S} = \b^\dagger_{1}\cdots\b^\dagger_{\nel}\ket{\vac} \,,\qquad   \b^\dagger_{j} = \sum_{k=1}^\nso Q_{jk}\ssp\c^\dagger_k\,,
\end{align}
where $Q$ is an $\nel\times \nso$ matrix satisfying
\begin{align}\label{eq:Q}
 Q^\dagger Q = P_S \,,
\end{align}
with $P_S$ the projector (of rank $\nel$) onto the subspace spanned by the single-particle wave functions of the occupied spin orbitals (rows of $Q$). An arbitrary Slater determinant~\eqref{eq:slater} can be prepared by applying a single-particle basis change $\mathcal U$ to an easy-to-prepare determinant in the computational basis~\eqref{eq:hatree_fock}:
\begin{align}\label{eq:slater_map}
 \ket{\Psi_S} = \mathcal U\ssp \c^\dagger_{1}\; \cdots\c^\dagger_{\nel}\ket{\vac} \,,\quad \mathcal U\ssp \c_j^\dagger\, \mathcal U^\dagger = \b_j^\dagger\,,
\end{align}
for $j=1,2,\ldots,\nel$. The unitary $\mathcal U$ corresponds to a fermionic Fourier transformation when the rows of $Q$ are orthonormal plane waves. The identity~\eqref{eq:Q} remains true under the transformation 
$ Q\rightarrow V\ssp Q$, where $V$ is an arbitrary $\nel\times \nel$ unitary matrix. Indeed, the Slater determinant~\eqref{eq:slater} remains the same (up to an overall phase) under the basis transformation $V$:
\begin{align}
 \bigg(\prod_{j=1}^{\nel} \sum_{k=1}^{\nel} V_{jk}\,\b^\dagger_k \bigg)\ket{\vac}
 %&= \det (V) \,\b^\dagger_{1}\cdots\b^\dagger_{\nel}\ket{\vac}\nonumber\\
 &= \det (V)\ssp \ket{\Psi_S}\,,\label{eq:det_unchange}
\end{align}
where $\det (V)$ is the determinant of $V$. The unitary $V$ can be chosen to be composed of a sequence of Givens rotations on neighboring rows of $Q$ that bring the matrix elements in its upper right corner to zeros. A Givens rotation takes the following form in the two relevant coordinate axes: 
\begin{align}\label{eq:givens_G}
 G(\theta, \varphi) &= \begin{pmatrix}
  \cos\theta  & -e^{i\varphi} \sin \theta \\[2pt]
  \sin \theta &  e^{i\varphi} \cos \theta  
 \end{pmatrix}\nonumber\\[3pt]
 &=\begin{pmatrix}
  \cos\theta  & - \sin \theta \\[2pt]
  \sin \theta &   \cos \theta  
 \end{pmatrix}
 \begin{pmatrix}
  1  & 0 \\
  0 &  e^{i\varphi} 
 \end{pmatrix}
 \,,
\end{align}
where we generalize the original definition by allowing complex matrix elements. For example, with $\nso=6$ and $\nel=3$, we can break $V$ into three Givens rotations,
\begin{align}
Q &\;\rightarrow\; \begin{pmatrix}
  \xx & \xx& \xx & \xx& \xx& {\azero 0}\\
  \xx & \xx&\xx & \xx& \xx & \xx \\
  \xx & \xx&\xx & \xx& \xx & \xx
 \end{pmatrix}
 \;\rightarrow\; \begin{pmatrix}
  \xx & \xx& \xx & \xx& \xx& 0\\
  \xx & \xx&\xx & \xx& \xx & {\azero 0} \\
  \xx & \xx&\xx & \xx& \xx & \xx
 \end{pmatrix}\nonumber\\[4pt]
 &\;\rightarrow\; \begin{pmatrix}
  \xx & \xx& \xx & \xx& {\azero 0}& 0\\
  \xx & \xx&\xx & \xx& \xx & 0 \\
  \xx & \xx&\xx & \xx& \xx & \xx
 \end{pmatrix}\;=\; VQ\,,
\end{align}
where $*$ represents an arbitrary matrix element, and the \azerotext\ matrix elements are zeroed out with Givens rotations. This procedure does not change the Slater determinant~\eqref{eq:det_unchange}, and no physical operation is required. However, the problem makes it clear that the number of physical Givens rotations can be reduced and how this can be achieved. 
 
\begin{figure}[htb]\label{fig:givens}
\centering
\includegraphics[width=0.41\textwidth]{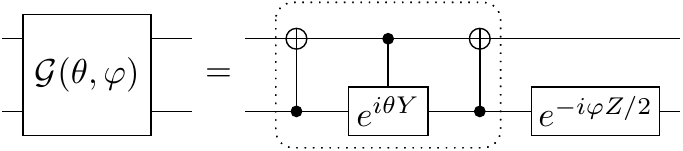}
\caption{Quantum circuit for a Givens rotation on neighboring qubits: the part in the dotted box represents a rotation between the two states $\ket{01}$ and $\ket{10}$.}
\end{figure}

The unitary in Eq.~\eqref{eq:slater_map} corresponds to a single-particle basis change and can be decomposed into a sequence of Givens rotations~\cite{wecker_solving_2015},
\begin{align}\label{eq:sU_G}
 \mathcal U = \mathcal G_1\ssp \mathcal G_2\nsp\cdots \mathcal G_{N_G}\,.
\end{align}
The Givens rotation $\mathcal G(\theta, \varphi)$ acting on the $j$th and $k$th spin orbitals takes the form
\begin{align}\label{eq:givens}
 \begin{pmatrix}
   \mathcal G\ssp\c_j^\dagger\ssp\mathcal G^\dagger \\[5pt] 
   \mathcal G\ssp\c_{k}^\dagger\ssp\mathcal G^\dagger 
 \end{pmatrix}
= 
G (\theta,\varphi)
 \begin{pmatrix}
   \c_j^\dagger  \\[5pt] \c_{k}^\dagger 
 \end{pmatrix}\,,
\end{align}
where the $2\times 2$ matrix $G(\theta,\varphi)$ is defined in Eq.~\eqref{eq:givens_G}. A Givens rotation of the form~\eqref{eq:givens} can be implemented on a quantum computer using the circuit in Fig.~\ref{fig:givens} (the two qubits are adjacent in the JWT). 
The first part of the circuit (the dotted box) describes a rotation between the qubit states $\ket{10}$ and $\ket{01}$ corresponding to a rotation in the single-particle subspace of the two spin orbitals; the two qubit states $\ket{00}$ and $\ket{11}$ corresponding to the unoccupied state and the double-occupied state are unchanged in this step. The fermionic unitary $\mathcal U$ in Eq.~\eqref{eq:sU_G} can be represented by a matrix $U$ corresponding to a single-particle basis change (see Appendix~\ref{sec:quadratic}),
\begin{align}\label{eq:U_G}
  \mathcal U\,\mathbf \c^\dagger\ssp\mathcal U^\dagger 
= U \mathbf \c^\dagger\,,\quad U = G_{N_G}\nsp \cdots G_2\ssp G_1 \,,
\end{align}
where $\mathbf{\c}^\dagger= (\begin{matrix}\c_{1}^\dagger & \cdots & \c_{\nso}^\dagger\end{matrix})^T$. The order of the Givens rotations is reversed in Eq.~\eqref{eq:U_G} compared to Eq.~\eqref{eq:sU_G}; this is the case because the matrix $G$ acts directly on the vector of fermionic creation operators in Eq.~\eqref{eq:givens}, and it should be placed on the right side of all matrices representing earlier rotations. To perform the transformation in Eq.~\eqref{eq:slater_map}, we require that the first $\nel$ rows of $U$ be equal to $VQ$; see also Appendix~\ref{sec:quadratic}. With the example $\nso=6$ and $\nel=3$, this condition is equivalent to
\begin{align}
V Q\ssp U^\dagger = 
 \begin{pmatrix}
  1 & 0& 0 & 0&0 &0\\
  0 & 1 &0 &0 &0 &0 \\
  0 & 0&1  &0 &0 &0
 \end{pmatrix}\,.
\end{align}
A sequence of Givens rotations---acting on adjacent columns of $VQ$ (corresponding to neighboring qubits in the JWT)---can achieve the desired form as follows:
\begin{align}\label{eq:VQ_VQUdag}
V Q
% \;=\;& 
% {\scriptstyle
% \begin{pmatrix}
%   \xx & \xx& \xx &\xx& 0& 0\\
%   \xx & \xx&\xx & \xx& \xx & 0 \\
%   \xx & \xx&\xx & \xx& \xx & \xx
%  \end{pmatrix}
%  }
 \,&\rightarrow\,
{
 \begin{pmatrix}
  \xx & \xx& \xx & {\azero 0}& 0& 0\\
  \xx & \xx&\xx & \xx& \xx & 0 \\
  \xx & \xx&\xx & \xx& \xx & \xx
 \end{pmatrix}}
 \,\rightarrow\,
 {
 \begin{pmatrix}
  \xx & \xx& {\azero 0} & 0& 0& 0\\
  \xx & \xx&\xx & \xx& {\azero 0} & 0 \\
  \xx & \xx&\xx & \xx& \xx & \xx
 \end{pmatrix}}\nonumber\\[5pt]
 &\rightarrow\,
 {
 \begin{pmatrix}
   {\pzero {\lambda_1}}\! & {\azero 0}& 0 & 0& 0& 0\\
  {\pzero 0}  & \xx&\xx & {\azero 0}& 0 & 0 \\
  {\pzero 0}  & \xx&\xx & \xx& \xx & {\azero 0}
 \end{pmatrix}}
 \,\rightarrow\,
 {
 \begin{pmatrix}
  \lambda_1\!& 0& 0 & 0& 0& 0\\
  0 &  {\pzero {\lambda_2}}\!&{\azero 0}& 0& 0 & 0 \\
  0 & {\pzero 0}&\xx & \xx& {\azero 0} & 0
 \end{pmatrix}}\nonumber\\[5pt]
 &\rightarrow\,
 {
 \begin{pmatrix}
  \lambda_1\! & 0& 0 & 0& 0& 0\\
  0 & \lambda_2\!&0& 0& 0 & 0 \\
  0 & 0& {\pzero {\lambda_3}}\! & {\azero 0}& 0 & 0
 \end{pmatrix}}
 \,\rightarrow \; V Q\ssp U^\dagger  \,,
\end{align}
where the $\lambda_j$ are phase factors, i.e., $\norm{\lambda_j}=1$. The \azerotext\ matrix elements are zeroed out in each step; Givens rotations on nonoverlapping columns can be implemented in parallel~\cite{kivlichan_quantum_2017}. The \pzerotext\ ones become zeros or phase factors automatically due to the orthonormality of the rows. The phase factors are brought to ones in the last step by single-qubit rotations; this step is unnecessary if the goal is to prepare a single Slater determinant. The total number of Givens rotations needed for the transformation $U$ is
\begin{align}\label{eq:n_gates}
 N_G &= \nso\nel -\nel(\nel-1)/2-\nel(\nel+1)/2\nonumber\\[3pt]
 &= (\nso-\nel)\nel\,,
\end{align}
and the circuit depth is $N-1$. Our algorithm requires only $N^2/4$ Givens rotations in the worst case, when $\nel=N/2$. By comparison, direct implementations without using the unitary freedom $V$ require more Givens rotations~\cite{wecker_solving_2015, kivlichan_quantum_2017}. We also point out that the trick to reducing $N_G$ by interchanging the roles of particles and holes~\cite{wecker_solving_2015} is no longer needed here; i.e., particles and holes are treated on an equal footing in our algorithm. 
% Our algorithm can be used to implement the 1D fermionic Fourier transformation on a qubit chain using $O(\nso^2)$ two-qubit gates with circuit depth $O(\nso)$.

In summary, we describe in this section a method to prepare a Slater determinant~\eqref{eq:slater} using two-qubit gates that act only on neighboring qubits. Our method can be achieved in four steps:
\begin{enumerate} 
 \item Store the wave functions of the occupied orbitals in the rows of the matrix $Q$.
 \item Zero out the upper-right matrix elements of $Q$ using the freedom $Q\rightarrow VQ$.
 \item\label{item:diagonalize} Diagonalize $VQ$ using a sequence of Givens rotations as column transformations. 
 \item Find the quantum gates that correspond to the Givens rotations in step~\ref{item:diagonalize}.
\end{enumerate}
As mentioned above, this algorithm is implemented as part of the open-source project  OpenFermion~\cite{mcclean_openfermion_2017}. The code
takes as input the matrix $Q$ from Eq.~\eqref{eq:slater} describing a Slater determinant and outputs a sequence of elements of the form $(j, k, \theta, \phi)$, which describes a Givens rotation of columns $j$ and $k$. Furthermore, rotations that can be performed in parallel are grouped together. We use the code to verify that the method described here does produce the desired Slater determinant.

\section{Preparing fermionic Gaussian states}
\label{sec:gaussian}

Fermionic Gaussian states~\cite{bach_generalized_1994,  greplova_quantum_2013} can be regarded as a generalization of Slater determinants obtained by relaxing the constraint that the total number of particles be fixed. The celebrated Bardeen-Cooper-Schrieffer (BCS) wave function~\cite{bardeen_microscopic_1957} for superconductivity is a special case of fermionic Gaussian states. Verstraete \textit{et al.}~\cite{verstraete_quantum_2009} demonstrated how to prepare the ground state of a BCS-like Hamiltonian---for 1D translationally invariant systems---using the fermionic fast Fourier transformation and the two-mode Bogoliubov transformation. Recently, superpositions of fermionic Gaussian states were used to approximate low-energy states of quantum impurity models~\cite{bravyi_complexity_2017}, which can be useful in a quantum-classical hybrid scheme for correlated materials~\cite{bauer_hybrid_2016}. Simulating quantum systems with disorder on a quantum computer may also lead to a better understanding of and, ultimately, control over the emergent phases from impurities~\cite{seo_disorder_2013}.

Here, we discuss how to prepare an arbitrary fermionic Gaussian state as the ground state of a quadratic Hamiltonian,
\begin{align}\label{eq:qua}
 \mathcal H = \sum_{j, k=1}^{\nso} \mm_{jk}\ssp \c_{j}^\dagger \c^{}_{k} + \half\sum_{j,k=1}^{\nso} \Big(\Delta_{jk}\ssp \c_{j}^\dagger \c_{k}^\dagger +\hc\Big)\,,
\end{align}
where $\mm = \mm^\dagger$ and $\Delta = - \Delta^T$ are complex matrices. \python\ code for this algorithm is available as part of the OpenFermion project~\cite{mcclean_openfermion_2017}. Our method can also be used to implement an arbitrary fermionic Gaussian unitary. In Appendix~\ref{sec:quadratic}, we review the standard approach to bringing the Hamiltonian~\eqref{eq:qua} into the diagonal form, 
\begin{align}
 \mathcal H = \sum_{j=1}^{\nso} \varepsilon_{j}\ssp\b_{j}^\dagger \b^{}_{j} + \cnumber\,, 
\end{align}
where $0\leq \varepsilon_{1} \leq \varepsilon_{2}\cdots  \leq \varepsilon_{\nso}$, and $\b_{j}$ and $\b_{j}^\dagger$ are a new set of fermionic operators that satisfy the canonical anticommutation relations. The new fermionic operators are linear combinations of the original ones:
\begin{align}
\begin{pmatrix}
 \mathbf{\b}^\dagger\\
 \mathbf{\b}
\end{pmatrix} 
 =  W
 \begin{pmatrix}
 \mathbf{\c}^\dagger\\
 \mathbf{\c} 
\end{pmatrix}= 
\begin{pmatrix}
 \mathcal W\,\mathbf{\c}^\dagger\ssp\mathcal W^\dagger\\
 \mathcal W\,\mathbf{\c}\,\mathcal W^\dagger
\end{pmatrix}\,,
\end{align}
where $(\begin{matrix}\mathbf{\c}^\dagger & \mathbf{\c} \end{matrix})^T= (\begin{matrix}\c_{1}^\dagger & \cdots & \c_{\nso}^\dagger & \c_{1}\cdots\, \c_{\nso}\end{matrix})^T$ and $(\begin{matrix}\mathbf{\b}^\dagger & \mathbf{\b}\end{matrix} )^T= (\begin{matrix}\b_{1}^\dagger & \cdots & \b_{\nso}^\dagger & \b_{1} & \cdots & \b_{\nso}\end{matrix})^T$. The fermionic Gaussian unitary $\mathcal W$  performs the linear transformation $W$, and the ground state of the quadratic Hamiltonian~\eqref{eq:qua} is
\begin{align}\label{eq:MF_ground}
 \ket{\Psi_0} = \mathcal W\, \ket{\vac} = \tran\, \ket{\vac} \,,
\end{align}
where $\c_j\ssp\ket{\vac} = 0$ for $j=1,2,\ldots,\nso$, and $\mathcal U = \mathcal W \mathcal V$ for some single-particle basis transformation $\mathcal V$. The unitary matrix $W$ has the block form
\begin{align}\label{eq:W}
W = 
\begin{pmatrix}
 W_1^* & W_2^*\\[2pt]
 W_2   & W_1
\end{pmatrix} \,, 
% \qquad W^\dagger W = 
% \begin{pmatrix}
%  \mathbb 1 & 0\\[2pt]
%  0   & \mathbb 1
% \end{pmatrix}\,,
\end{align}
where the submatrices satisfy
\begin{gather}\label{eq:W_a}
W_1 W_1^\dagger + W_2 W_2^\dagger = \mathbb 1\,,\\[2pt]
W_1 W_2^T + W_2 W_1^T = \mathbb 0\,,\label{eq:W_b}
\end{gather}
with $\mathbb 1$ and $\mathbb 0$ being the $\nso\times \nso$ identify matrix and the zero matrix, respectively. We define the $\nso\times 2\nso$ matrix $W_L = (\begin{matrix} W_2 & W_1\end{matrix})$ as the lower half of $W$; the $j$th row of $W_L$ corresponds to the expansion coefficients of the operator $\b_j$. The matrix $W_L$ uniquely determines the transformation $\mathcal W$ up to an overall phase.

Using elementary matrix manipulations on $W_L$, we demonstrate that the Gaussian unitary $\tran$ in Eq.~\eqref{eq:MF_ground} can be broken into a sequence of operations that can be implemented on a quantum computer:
\begin{align}
 \tran =  \mathcal \ph\ssp \mathcal G_1 \mathcal \ph\ssp\mathcal G_2 \ssp\mathcal G_3\mathcal \ph  \cdots \mathcal G_{N_G}\ssp\mathcal\ph\,,
\end{align}
where the $\mathcal G_j$ are Givens rotations on adjacent fermionic modes encoded in the JWT, and $\mathcal \ph$ denotes the particle-hole Bogoliubov transformation on the last fermionic mode,
\begin{align}
&\mathcal \ph\ssp \c_\nso \mathcal \ph^\dagger = \c_\nso^\dagger\,,\\
&\mathcal \ph\ssp \c_j\ssp \mathcal \ph^\dagger = \c_j,\; \text{for $j=1,2,\ldots, \nso-1$}\,.
\end{align}
The transformation $\mathcal \ph$ does not conserve the total number of particles and is crucial to preparing superpositions of states with different numbers of particles. It can be implemented easily by applying the Pauli-$X$ operator on the last qubit; the parities of the other fermionic modes encoded in the JWT are not affected, which would not be true for any other spin orbital. In the relevant coordinate axes, the matrix representation of the Givens rotation is 
\begin{align}\label{eq:G}
G = \left(
\begin{array}{cc|cc}
  \cos\theta  & -e^{i\varphi} \sin \theta & 0& 0\\
  \sin \theta &  e^{i\varphi} \cos \theta   & 0& 0\\
  \hline
  0& 0 &\cos\theta  & -e^{-i\varphi} \sin \theta \\
  0& 0 &\sin \theta &  e^{-i\varphi} \cos \theta  
\end{array}
\right)\,.
\end{align}
The representation of the particle-hole transformation is
\begin{align}\label{eq:F}
\ph = \ph^\dagger =
\begin{pmatrix}
  \mathbb 1-e_\nso e_\nso^T & e_\nso e_\nso^T \\[4pt]
  e_\nso e_\nso^T & \mathbb 1-e_\nso e_\nso^T
\end{pmatrix}\,,
\end{align}
where $e_\nso  = (\begin{matrix}0&0&\cdots& 0&1 \end{matrix})^T$ is an $\nso$-dimensional unit vector. The goal is to find a $2\nso\times 2\nso$ unitary matrix $U$ decomposed into $G$ and $\ph$,
\begin{align}
 U = \ph\ssp G_{N_G}\cdots \ph\ssp G_3 \ssp G_2\ph\ssp G_1\ssp \ph\,,
\end{align}
such that
\begin{align}\label{eq:0V}
V W_L U^\dagger = 
\big(
\begin{array}{cc}
\mathbb 0 & \mathbb 1
\end{array}
\big)\,,
\end{align}
where $V$ is an arbitrary unitary matrix. The right-hand side of Eq.~\eqref{eq:0V} represents the original annihilation operators $\c_j$, which corresponds to the vacuum state defined in Eq.~\eqref{eq:MF_ground}. We discuss how to bring $W_L$ to the desired form~\eqref{eq:0V} using an example of four spin orbitals. Following Sec.~\ref{sec:slater}, we use the unitary $V$ as a freedom to zero out some matrix elements on the left side of $W_L$,
\begin{align}\label{eq:V_WL}
V W_L = 
{
\left(
\begin{array}{cccc|cccc}
0 & 0 & 0 &\xx\; & \;\xx & \xx & \xx & {\pzero 0} \\[-2pt]
0 & 0 & \xx &\xx\; & \;\xx & \xx & \xx & \xx \\[-2pt]
0 & \xx & \xx &\xx\; & \;\xx & \xx & \xx & \xx \\[-2pt]
\xx & \xx & \xx &\xx\; & \;\xx & \xx & \xx & \xx 
\end{array}
\right)}
\,,
\end{align}
where $\xx$ represents an arbitrary matrix element and the \pzerotext\ matrix element is automatically zeroed out due to Eq.~\eqref{eq:W_b}. A sequence of elementary operations that brings $W_L$ to the desired form is
\begin{widetext}
 \begin{align}\label{eq:V_WL_Udag}
\begin{split}
V W_L %\;=\; 
% {\scriptscriptstyle
% \left(
% \begin{array}{cccc|cccc}
% 0 & 0 & 0 &\xx\; & \;\xx & \xx & \xx & {\pzero 0} \\[-2pt]
% 0 & 0 & \xx &\xx\; & \;\xx & \xx & \xx & \xx \\[-2pt]
% 0 & \xx & \xx &\xx\; & \;\xx & \xx & \xx & \xx \\[-2pt]
% \xx & \xx & \xx &\xx\; & \;\xx & \xx & \xx & \xx 
% \end{array}
% \right)}
\,&\rightarrow
{\scriptscriptstyle
\left(
\begin{array}{cccc|cccc}
0 & 0 & 0 &{\azero 0}\; & \;\xx & \xx & \xx & {\azero \xx}\\[-2pt]
0 &   0 & \xx & \xx \; & \;\xx & \xx & \xx & \xx \\[-2pt]
0 & \xx & \xx &\xx\; & \;\xx & \xx & \xx & \xx \\[-2pt]
\xx & \xx & \xx &\xx\; & \;\xx & \xx & \xx & \xx 
\end{array}
\right)}
\rightarrow
{\scriptscriptstyle
\left(
\begin{array}{cccc|cccc}
0 & 0 & 0 &0\; & \;\xx & \xx & \xx & {\pzero 0}\\[-2pt]
0 & 0 & {\azero 0}  &\xx\; & \;\xx & \xx & \xx & {\pzero 0} \\[-2pt]
0 & \xx & \xx &\xx\; & \;\xx & \xx & \xx & \xx \\[-2pt]
\xx & \xx & \xx &\xx\; & \;\xx & \xx & \xx & \xx
\end{array}
\right)}
\rightarrow
{\scriptscriptstyle
\left(
\begin{array}{cccc|cccc}
0 & 0 & 0 &0\; & \;\xx & \xx & {\pzero 0} & 0\\[-2pt]
0 & 0 & 0  &{\azero 0}\; & \;\xx & \xx & \xx & {\azero \xx} \\[-2pt]
0 & {\azero 0} & \xx &\xx\; & \;\xx & \xx & \xx & \xx \\[-2pt]
\xx & \xx & \xx &\xx\; & \;\xx & \xx & \xx & \xx
\end{array}
\right)}
\rightarrow
{\scriptscriptstyle
\left(
\begin{array}{cccc|cccc}
0 & 0 & 0 &0\; & \;{\pzero {\lambda_1}\!} & {\pzero 0} & 0 & 0\\[-2pt]
0 & 0 & 0  &0\; & \;{\pzero 0} & \xx & \xx & {\pzero 0} \\[-2pt]
0 & 0 & {\azero 0} &\xx\; & \;{\pzero 0} & \xx & \xx & {\pzero 0}\\[-2pt]
{\azero 0} & \xx & \xx &\xx\; & \;{\pzero 0} & \xx & \xx & \xx
\end{array}
\right)}\\[6pt]
&\rightarrow
{\scriptscriptstyle
\left(
\begin{array}{cccc|cccc}
0 & 0 & 0 &0\; & \;\lambda_1\! & 0 & 0 & 0 \\[-2pt]
0 & 0 & 0 &0\; & \;0 & {\pzero {\lambda_2}\!} & {\pzero 0} & 0 \\[-2pt]
0 & 0 & 0 &{\azero 0}\; & \;0  & {\pzero 0} & \xx & {\azero \xx} \\[-2pt]
0 & {\azero 0} & \xx &\xx\; & \;0 & {\pzero 0} & \xx & \xx
\end{array}
\right)}
\rightarrow
{\scriptscriptstyle
\left(
\begin{array}{cccc|cccc}
0 & 0 & 0 &0\; & \;\lambda_1\! & 0 & 0 & 0 \\[-2pt]
0 & 0 & 0 &0\; & \;0 & \lambda_2\! & 0 & 0 \\[-2pt]
0 & 0 & 0 &0\; & \;0 & 0 & {\pzero {\lambda_3}\!} & {\pzero 0} \\[-2pt]
0 & 0  & {\azero 0} &{\pzero {\lambda_4}\!}\; & \;0 & 0 &{\pzero 0}& {\pzero 0} 
\end{array}
\right)}
\rightarrow
{\scriptscriptstyle
\left(
\begin{array}{cccc|cccc}
0 & 0 & 0 &0\; & \;\lambda_1\! & 0 & 0 & 0 \\[-2pt]
0 & 0 & 0 &0\; & \;0 & \lambda_2\! & 0 & 0 \\[-2pt]
0 & 0 & 0 &0\; & \;0 & 0 & \lambda_3\! & 0 \\[-2pt]
0 & 0 & 0 &{\azero 0}\; & \;0 & 0 & 0 & {\azero {\lambda_4}\!}
\end{array}
\right)}
\rightarrow\, V W_L U^\dagger\,.
\end{split}
\end{align}
\end{widetext}
The \azerotext\ matrix elements in the fourth column are always zeroed out by the particle-hole transformation $\ph$, and the other \azerotext\ matrix elements on the left side are zeroed out by the Givens rotations $G$. The \azerotext\ elements on the right side become nonzero due to the particle-hole transformation $\ph$, and the \pzerotext\ matrix elements are brought to zeros or phase factors automatically by the condition~\eqref{eq:W_a} or \eqref{eq:W_b}. The phase factors are brought to ones in the last step by single-qubit rotations; this step is unnecessary if $\mathcal U$ is applied to the vacuum state. The total numbers of Givens rotations and particle-hole transformations are
\begin{align}
 N_G = (\nso-1)\nso/2\,,\qquad N_\ph = \nso\,,
\end{align}
and the circuit depth is, at most, $2\nso-1$.

In summary, we describe in this section a method to prepare an arbitrary Gaussian state as the ground state of the quadratic Hamiltonian~\eqref{eq:qua} using two-qubit gates that act only on neighboring qubits. Our method can be achieved in four steps:
\begin{enumerate}
 \item Calculate the matrix $W$ using the procedure described in Appendix~\ref{sec:quadratic}.
 \item Zero out the upper-left matrix elements of $W_L$ using the freedom $W_L\rightarrow VW_L$~\eqref{eq:V_WL}.
 \item \label{item:sequence} Zero out the remaining matrix elements on the left side of $VW_L$ using the sequence~\eqref{eq:V_WL_Udag}.
 \item Find the quantum gates corresponding to the sequence in step~\ref{item:sequence}.
\end{enumerate}

The procedure described here also applies to the implementation of an arbitrary Gaussian unitary, where one cannot use the unitary freedom $V$. By rearranging Eq.~\eqref{eq:0V}, we have
\begin{align}
 W_L  = V^\dagger \,
\big(
\begin{array}{cc}
\mathbb 0 & \mathbb 1
\end{array}
\big)\ssp U = 
\big(
\begin{array}{cc}
\mathbb 0 & \mathbb 1
\end{array}
\big) \bigg(
\begin{array}{cc}
V^T & \mathbb 0\\ \mathbb 0 & V^\dagger
\end{array}
\bigg)\, U\,,
\end{align}
where the matrix $\diag \big(V^T,\, V^\dagger\big)$ corresponds to a basis change and can be decomposed as Givens rotations using the method discussed in Sec.~\ref{sec:slater}.

As mentioned above, this algorithm has also been implemented as part of OpenFermion~\cite{mcclean_openfermion_2017}.
The user can specify a quadratic Hamiltonian by inputting the matrices $M$ and $\Delta$ from Eq.~\eqref{eq:qua}. The code then outputs a sequence of operations, each of which either describes a Givens rotation or indicates a particle-hole transformation on the last fermionic mode. The operations that can be performed in parallel are grouped together. Since OpenFermion also contains modules to initialize common Hamiltonian models, it can be used to easily specify, for instance, a mean-field Hamiltonian and then obtain a circuit that prepares its ground state. We use the code to verify that the method described here does produce the desired ground state.

\section{Fermionic Fourier transformations}
\label{sec:fourier}

The fermionic Fourier transformation is a subroutine of many quantum algorithms. It was first introduced for quantum-computing purposes in Ref.~\cite{verstraete_quantum_2009}. Recently, Babbush \ea~\cite{babbush_low_2017} demonstrated that the 2D and 3D fermionic Fourier transformations can be implemented using a 2D qubit array with $O(\nso)$ depth. Here, we present an algorithm to implement the 2D fermionic Fourier transformation using a 2D qubit array with only $O(\sqrt\nso)$ depth. This amount of depth is required for quantum information to travel across the array, and therefore this scaling is optimal. Our method provides an example where the parity problem in JWTs for more than one spatial dimension can be circumvented with negligible overhead. Our algorithm also works for more general transformations that can be factorized into a product of transforms on each spatial dimension, including Fourier transformations with open-boundary conditions and fermionic Gaussian unitaries. This algorithm allows one to efficiently prepare the initial states for systems whose ground states are well approximated by the mean-field states, as well as to improve the efficiency of measurements.

\begin{figure}[ht]\label{fig:snake}
\centering
\includegraphics[width=0.23\textwidth]{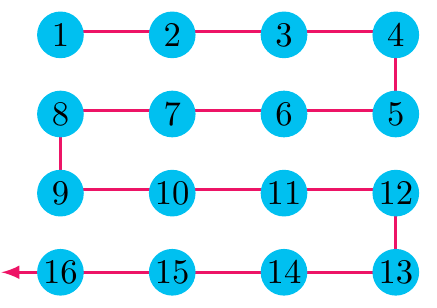}
\caption{The fermionic spin orbitals on a $4\times 4$ array are mapped to qubits (the blue circles) on an array of the same dimensions using a snake-shaped JWT. The red arrow shows the direction of the JWT, and the qubits are numbered by their order in the JWT. Hopping terms between an odd-numbered row and the row below are called right-closed hopping terms, while those between an even-numbered row and the row below are called left-closed hopping terms.}
\end{figure}

In Sec.~\ref{sec:slater}, we demonstrate that any fermionic single-particle basis transformation---including the 1D fermionic Fourier transformation---can be implemented using a 1D qubit chain with $O(\nso^2)$ gates and $O(\nso)$ depth; see also Theorem~7 in Ref.~\cite{babbush_low_2017}. Using this algorithm as a subroutine, we discuss an algorithm that implements the 2D fermionic Fourier transformation using a 2D qubit array with $O(\nso^{1.5})$ gates and $O(\sqrt\nso\ssp\ssp)$ depth. More generally, our method works for any fermionic single-particle basis transformation $\mathcal F$ (or fermionic Gaussian unitary) that factorizes,  
\begin{align}\label{eq:fac_trans}
 \mathcal F = \mathcal F_x \mathcal F_y = \mathcal F_y \mathcal F_x\,,
\end{align}
where the horizontal (vertical) transformation $\mathcal F_x$ ($\mathcal F_y$) is a product of commuting terms, each of which involves spin orbitals only in the same row (column). We map the fermionic operators to qubit operators using a snake-shaped JWT in row-major order; see Fig.~\ref{fig:snake}.
The transformation $\mathcal F_x$ on a single row can be implemented with $O(\nn_x^2)$ gates and $O(\nn_x)$ depth using the algorithm described in Sec.~\ref{sec:slater}, where $\nn_x$ is the number of spin orbitals in a single row (number of columns). The vertical transformation $\mathcal F_y$ is much harder to implement with our mapping because of the nonlocal parity operators in the hopping terms; see Eq.~\eqref{eq:hopping}. This difficulty can be overcome by reordering the fermionic spin orbitals in the JWT using the fermionic \SWAP\ gates; such a strategy allows one to perform the 2D fermionic Fourier transformation using a 2D qubit array with $O(\nso^{2})$ gates and $O(\nso)$ depth~\cite{babbush_low_2017}. 

We use a different approach that takes full advantage of the 2D qubit interactions. In our scheme, the transformation~\eqref{eq:fac_trans} is realized using the decomposition
\begin{align}\label{eq:F_decomp}
 \mathcal F = \mathcal F_x \mathcal F_y = \mathcal F_x\ssp \Gamma^\dagger \bF{y}\, \Gamma\,,
\end{align}
where $\Gamma = \Gamma^\dagger$ is a diagonal unitary matrix in the computational (Pauli-$Z$) basis with eigenvalues $\pm 1$. We will show that $\Gamma$ can be implemented with $O(\nso)$ gates and $O(\sqrt\nso\ssp\ssp)$ depth. The unitary $\bF{y}$ is implemented by using the Givens rotations without the parity operators attached, see Fig.~\ref{fig:givens}. This is equivalent to using a JWT with column-major order, and the generator of the real Givens rotation takes the form~\eqref{eq:n_hopping_a}
\begin{align}\label{eq:K}
 \bK_{jk} = \frac{1}{2}\ssp \big(X_j Y_k - Y_j X_k\big)\,,
\end{align}
where the qubits $j$ and $k$ are adjacent in one column; we call these operators bare hopping terms. The unitary transformation $\Gamma$ attaches the corresponding parity operator to the bare hopping terms,
\begin{align}\label{eq:dressing}
 \Gamma^\dagger \bK_{jk}\Gamma = \bK_{jk}\ssp Z_{j+1}\cdots Z_{k-1}\,,\quad k>j\,,
\end{align}
for any indices $j$ and $k$ adjacent in the same column. Any non-neighboring vertical hopping terms can be derived as (nested) commutators of the nearest-neighbor ones, and their parities are taken care of automatically. Because $\mathcal F_y$ can be decomposed into a product of Givens rotations generated by the vertical hopping terms, we have $\Gamma^\dagger \bF{y}\, \Gamma = \mathcal F_y$. We denote an arbitrary state in the computational basis with $\ket{\sysbf}$, where $\sysbf = (\sys_1,\sys_2,\ldots, \sys_\nso)$ is a binary string. The matrix element of the hopping term $\bK_{jk}$ with respect to the basis states $\ket{\sysbf}$ and $\ket{\sysbf'}$ takes the form 
\begin{align} \label{eq:parity_K}                                                                                                                                                                                                                                                                                                                                                                                                                                                                                                                                                                                                                                    \bra{\sysbf^\prime}\ssp\Gamma^\dagger\bK_{jk}\Gamma\ssp\ket{\sysbf}= \gamma_{\sysbf}\ssp \gamma_{\sysbf^\prime}\, \bra{\sysbf^\prime}\bK_{jk}\ket{\sysbf} \,,
\end{align}
where $\gamma_{\sysbf}$ and $\gamma_{\sysbf'}$ are the corresponding eigenvalues of $\Gamma$. 

Comparing Eq.~\eqref{eq:dressing} to Eq.~\eqref{eq:parity_K}, we have
\begin{align}
 \gamma_{\sysbf} \gamma_{\sysbf^\prime} = (-1)^{\sum_{l=j+1}^{k-1}\sys_l }=(-1)^{\sum_{l=j+1}^{k-1}\sys_l^\prime }\,,
\end{align}
for any pair of basis states such that $\bra{\sysbf^\prime}\bK_{jk}\ket{\sysbf}\neq 0$. The matrix element $\bra{\sysbf^\prime}\bK_{jk}\ket{\sysbf}$ is nonzero only when $\sys_j \neq \sys_j^\prime$, $\sys_k \neq \sys_k^\prime$, and $\sys_l = \sys_l^\prime$ for all $l\neq j,k$; the total parity of the qubits $j$ and $k$ also needs to be odd,
\begin{align}\label{eq:odd_parity}
 \sys_j + \sys_k = \sys_j^\prime +\sys_k^\prime = 1\,.
\end{align}
The unitary $\bF{y}$ on a single column can be implemented with $O(\nn_y^2)$ gates and $O(\nn_y)$ depth using the bare hopping terms, where $\nn_y$ is the number of spin orbitals per column. By parallelizing operations on different rows or columns of qubits, one can implement either $\mathcal F_x$ or $\bF{y}$ with $O(\nso^{1.5})$ gates and $O(\sqrt\nso\ssp\ssp)$ depth, where $\nso=\nn_x\times\nn_y$ is the number of spin orbitals. A Slater determinant in the momentum basis can be prepared by applying the transformation $\mathcal F$ to a Slater determinant in the site basis $\ket{\sysbf}$,
\begin{align}
 \mathcal F\ssp \ket{\sysbf} = \mathcal F_x\ssp \Gamma^\dagger \bF{y}\ssp \Gamma \ket{\sysbf} = \gamma_\sysbf \,\mathcal F_x\ssp \Gamma^\dagger \bF{y} \ket{\sysbf}\,,
\end{align}
where $\gamma_\sysbf=\pm 1$. The transformation $\Gamma$ is also useful for simulating the time evolution of fermionic systems, e.g., the Fermi-Hubbard model. The hopping terms in each Trotter step can be implemented with $O(\nso)$ gates and $O(\sqrt \nso)$ circuit depth by using $\Gamma$. 

\begin{figure}[htb]\label{fig:swcn}
\centering
\includegraphics[height=1.3cm]{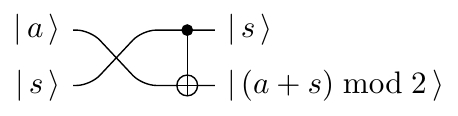} 
\caption{The ancilla qubit $\anc$ is swapped with the system qubit $\sys$ to its left, and its state is then updated by the \CNOT\ gate.}
\end{figure}

\begin{figure}[htb]\label{fig:right-closed}
\centering
\subfloat[]{\label{fig:right-closed_a}
\includegraphics[height=2.7cm]{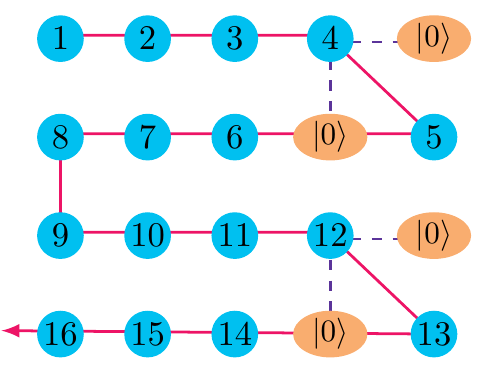} %f2
}\hspace{2pt} 
\subfloat[]{\label{fig:right-closed_b}
\includegraphics[height=2.7cm]{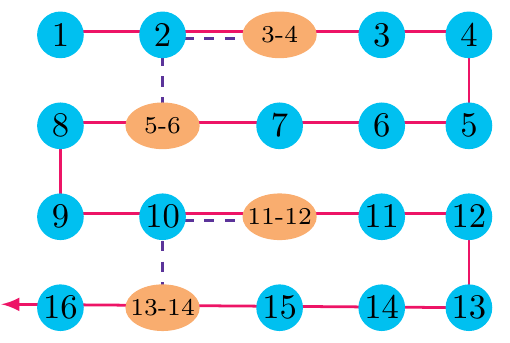}
}
\caption{The procedure to generate the desired parities for the right-closed hopping terms. The ancilla qubits (the orange ellipses) store the parity of the system qubits (the blue circles), where $j\myhyphen k$ denotes the parity $\big(\sum_{l=j}^{k}\sys_l\big)\! \mod 2$. \protect\subref{fig:right-closed_a} The ancilla qubits are initially located at the right side of the lattice, with their states set to $\ket{0}$. \protect\subref{fig:right-closed_b} They are then moved to the left while the parity stored in them being updated. The purple dashed lines represent \CZ\ gates between the ancilla and the system qubits for right-closed hopping terms. The \CZ\ gates in \protect\subref{fig:right-closed_a} can be omitted because the ancilla states are set to $\ket{0}$ initially.}
\end{figure}

To implement $\Gamma$, we introduce one ancilla qubit per row to store the parities (see Fig.~\ref{fig:right-closed}); they are used only to facilitate the implementation of $\Gamma$ and are disentangled with the system qubits in the end. Initially, the ancilla qubits are located on the right side in all of the system qubits, and their states are set to $\ket{0}$. In each time step, each ancilla qubit is swapped with the system qubit to the left, which allows it to interact with other system qubits. The parity stored in the ancilla qubit is updated by applying the \CNOT\ gate controlled by the same system qubit (now on the right side of the ancilla); see Fig.~\ref{fig:swcn}.
After the \CNOT\ gate, the ancilla qubits store the total parity of system qubits to their right on the same rows. We apply \CZ\ gates between the ancilla qubits and the system qubits (see Fig.~\ref{fig:right-closed}); each \CZ\ gate introduces a parity (an overall $\pm 1$ sign) to the state $\ket{\sysbf}$ in the computational basis. The goal is to find the set of \CZ\ gates, acting on neighboring qubits, such that they put the desired parities with the bare vertical hopping terms. It is instructive to first work out the cases for right-closed and left-closed hopping terms separately.

\begin{figure}[htb]\label{fig:cz_identity_a}
\centering
\includegraphics[height=2cm]{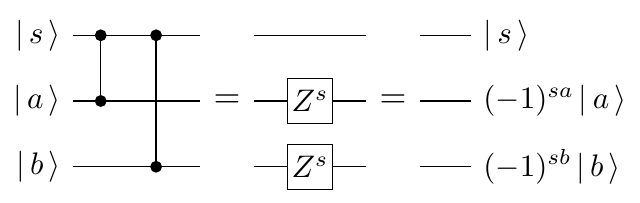} 
\caption{The \CZ\ gates involving one system qubit $\sys$ and two ancilla qubits $\anc$ and $\bnc$. When $\sys\neq \sys'$ and $\anc+ \bnc = \anc'+ \bnc'$ for the two basis states, a parity difference of $\anc+ \bnc$ is introduced. By comparison, no parity difference is introduced when both quantities are the same for the two basis states.}
\end{figure}

In Fig.~\ref{fig:right-closed}, we plot the \CZ\ gates (the dashed lines) that generate the desired parities for the right-closed hopping terms between an odd-numbered row and the row below it. We explain how this works for the hopping term between the system qubits 2 and 7.
Any two basis states $\ket{\sysbf}$ and $\ket{\sysbf^\prime}$ corresponding to a nonzero matrix element of the hopping term satisfy $\sys_l =\sys_l^\prime$ for $l\in \{3,4,5,6\}$; therefore, they accumulate the same parity before hitting the two \CZ\ gates involving qubit 2; see Fig.~\subref*{fig:right-closed_b}. The states of the two ancilla qubits involved in these two \CZ\ gates are the same for the two basis states because the parities of qubits 2 and 7 have not been added to them. After the two \CZ\ gates in Fig.~\subref*{fig:right-closed_b}, the two basis states acquire a difference in parity equaling $(\sys_3+\sys_4+\sys_5+\sys_6)\!\! \mod 2$; see Fig.~\ref{fig:cz_identity_a}. To simplify the notations, we will neglect mod 2 in dealing with parities hereafter. 
The two ancilla qubits in the first and second rows take different values for $\ket{\sysbf}$ and $\ket{\sysbf^\prime}$ after the parities of qubits 2 and 7 are added to them; however, the total parities of the two ancilla qubits are still the same. The subsequent two \CZ\ gates (qubit 1 is involved) do not introduce a parity difference to the two basis states because both $\sys$ and $\anc+ \bnc$ are the same for the two basis states. After the ancilla qubits reach the left end, all of the right-closed hopping terms get the desired parities from the \CZ\ gates. 

\begin{figure}[htb]\label{fig:left-closed}
\centering
\subfloat[]{\label{fig:left-closed_a}
\includegraphics[height=2.7cm]{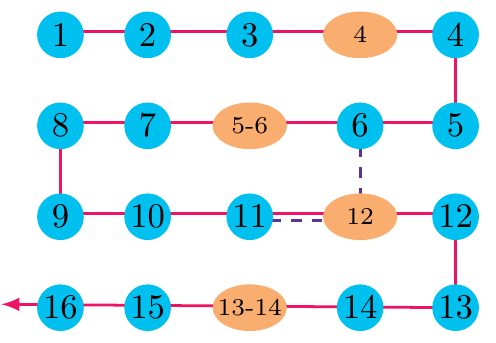} %f2
}\hspace{2pt} 
\subfloat[]{\label{fig:left-closed_b}
\includegraphics[height=2.7cm]{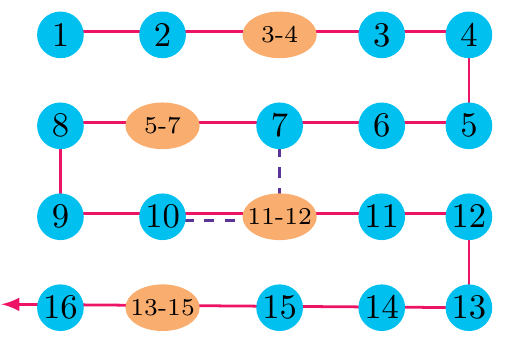}
}
\caption{The procedure to generate the desired parities for the left-closed hopping terms. The ancilla qubits (the orange ellipses) store the parity of the system qubits (the blue circles), where $j\myhyphen k$ denotes the parity $\big(\sum_{l=j}^{k}\sys_l\big)\! \mod 2$. \protect\subref{fig:left-closed_a} Two \CZ\ gates (the dashed lines) are applied to the anilla qubit labeled 12 and the system qubits labeled 6 and 11. \protect\subref{fig:left-closed_b} Two \CZ\ gates are applied to the anilla qubit labeled $11\myhyphen 12$ and the system qubits labeled 7 and 10.}
\end{figure}

\begin{figure}[htb]\label{fig:cz_identity_b}
\centering
\includegraphics[height=2cm]{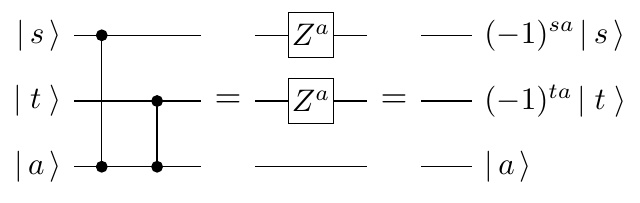} 
\caption{The \CZ\ gates involving two system qubits $\sys$ and $\tys$ and one ancilla qubit $\anc$. When $\anc\neq \anc'$ and $\sys+ \tys = \sys'+ \tys'$ for the two basis states, a parity difference of $\sys+ \tys$ is introduced. By comparison, no parity difference is introduced when both quantities are the same.}
\end{figure}

In Fig.~\ref{fig:left-closed}, we plot the \CZ\ gates that introduce the desired parities for the left-closed hopping terms, i.e., those between an even-numbered row and the row below.
We now explain how this works for the hopping term between qubits 6 and 11. Any two basis states $\ket{\sysbf}$ and $\ket{\sysbf^\prime}$ corresponding to a nonzero matrix element of the hopping term satisfy $\sys_{l}=\sys_{l}^\prime$ for $l\in\{5,12\}$; therefore, they accumulate the same parities before hitting the two \CZ\ gates in Fig.~\subref*{fig:left-closed_a}. These two \CZ\ gates do not introduce a parity difference, either, because both the parity of the ancilla qubit and the total parity of the two system qubits are the same for the two basis states; see Eq.~\eqref{eq:odd_parity} and Fig.~\ref{fig:cz_identity_b}.
The ancilla qubit takes different values for $\ket{\sysbf}$ and $\ket{\sysbf^\prime}$ after the parity of qubit 11 is added to it. As a result, the following two \CZ\ gates (qubits 7 and 10 are involved) introduce a parity difference to the two basis states equaling $\sys_7 +\sys_{10}= \sys_7^\prime+ \sys_{10}^\prime$; see Fig.~\ref{fig:cz_identity_b}. Since the state of the ancilla qubit remains different for $\ket{\sysbf}$ and $\ket{\sysbf^\prime}$, a parity difference of $\sys_8 +\sys_{9}= \sys_8^\prime +\sys_{9}^\prime$ is introduced in the next step. After the ancilla qubits reach the left end, all of the left-closed hopping terms get the desired parities from the \CZ\ gates. 
 
\begin{figure}[htb]\label{fig:single_hop}
\subfloat[]{\label{fig:single_hop_a}
\includegraphics[width=0.34\linewidth]{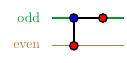}}\hspace{0.6cm}
\subfloat[]{\label{fig:single_hop_b}
\includegraphics[width=0.34\linewidth]{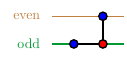}}
\caption{The \CZ\ gates between system (blue) and ancilla (red) qubits. (a) Right-closed hopping terms (between an odd-numbered row and the row below it), e.g., the one involving qubit 2 in Fig.~\protect\subref*{fig:right-closed_b}. (b) Left-closed hopping terms (between an even-numbered row and the row below it), e.g., the one involving qubits 6 and 11 in Fig.~\protect\subref*{fig:left-closed_a}
}
\end{figure}

We have shown how to introduce the desired parities to the right-closed and the left-closed hopping terms using the \CZ\ gates in Figs.~\ref{fig:right-closed} and~\ref{fig:left-closed}, respectively. The ancilla qubits are brought in to interact with system qubits in a particular column at a time, and they store the total parities of all of the system qubits in the same row to the right of the current column. We depict the two kinds of \CZ\ gates in Fig.~\ref{fig:single_hop} for a single time step, where the system (ancilla) qubits are represented by blue (red) dots. 
One will not achieve the desired results for both kinds of hopping terms by simply combining the two sets of \CZ\ gates. This is the case because the \CZ\ gates for the right-closed terms also affect the parities of the left-closed terms, and vice versa. 

\begin{figure}[htb]\label{fig:many_hop_a}
\includegraphics[height=3.2cm]{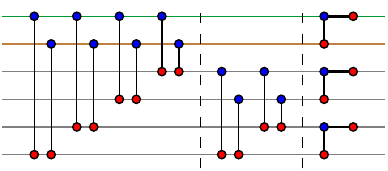}
\caption{The \CZ\ gates between the system (blue) and ancilla (red) qubits in a single time step that introduce the desired parities to both the right-closed and left-closed hopping terms.
}
\end{figure}

To circumvent this difficulty, we introduce the \CZ\ gates for a single time step using an example of 6 rows in Fig.~\ref{fig:many_hop_a}. 
In addition to the \CZ\ gates in Fig.~\ref{fig:single_hop} for all of the right-closed hopping terms, we also introduce a \CZ\ gate between each system qubit and the ancilla qubits below it, starting from the next odd-numbered row. We show that this set of \CZ\ gates works for both right-closed and left-closed hopping terms. First, we discuss the hopping term between the first and second rows. The \CZ\ gates are divided into three parts in Fig.~\ref{fig:many_hop_a}. In the first part, the states of the ancilla qubits are the same for the two basis states $\ket{\sysbf}$ and $\ket{\sysbf^\prime}$, and the total parity of the two system qubits in the first and second rows are also the same; see Eq.~\eqref{eq:odd_parity}. As a result, the \CZ\ gates in the first part do not introduce a parity difference to the two basis states. Neither do the \CZ\ gates in the second part, because they act on qubits that are not involved in the hopping term. The third part is equivalent to the circuit in Fig.~\subref*{fig:single_hop_a}. Therefore, this circuit works for any hopping term between the first and second rows. 

\begin{figure}[htb]\label{fig:many_hop_b}
\includegraphics[height=3.2cm]{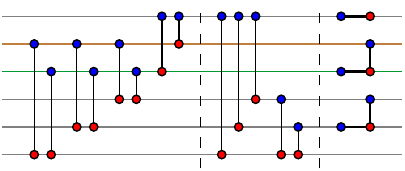}
\caption{Quantum circuit to demonstrate the hopping terms between the second and third rows by rearranging the \CZ\ gates in Fig.~\ref{fig:many_hop_a}}
\end{figure}

To show how the same circuit works for the hopping terms between the second and third rows, we rearrange the \CZ\ gates in Fig.~\ref{fig:many_hop_a} into the specific form in Fig.~\ref{fig:many_hop_b}. Because all \CZ\ gates commute, these two circuits are equivalent. 
The first three pairs of \CZ\ gates and the second part in Fig.~\ref{fig:many_hop_b} do not introduce a parity difference to the two basis states, for the same reasons we discussed earlier. Neither do the last two \CZ\ gates in the first part because the total parity of the two ancilla qubits in the second and third rows are the same for the two basis states. The third part of Fig.~\ref{fig:many_hop_b} is equivalent to the circuit in Fig.~\subref*{fig:single_hop_b}. Therefore, the same circuit also works for any hopping term between the second and third rows. It is straightforward to verify that the circuit works for any other nearest-neighbor vertical hopping terms in this example; see Appendix~\ref{sec:circuits_Gamma}. Therein, we also give an argument on why our approach works for systems with any even number of rows (an unused row can be added when the number of rows is odd).

\begin{figure}[htb]\label{fig:cz_s2_a}
\includegraphics[height=3.2cm]{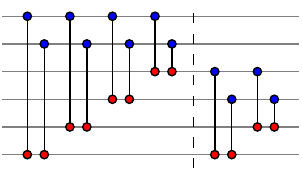}
\caption{The first and second parts in Fig.~\ref{fig:many_hop_a}}
\end{figure}

\begin{figure}[htb]\label{fig:parity_basis_a}
\subfloat[]{\label{fig:add_parity}
\includegraphics[height=3.1cm]{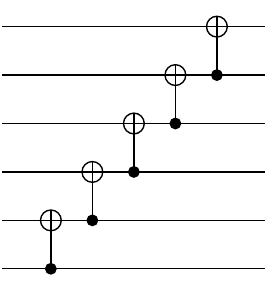}}\hspace{0.6cm}
\subfloat[]{\label{fig:parity_example}
\raisebox{0.3cm}{\includegraphics[height=2.7cm]{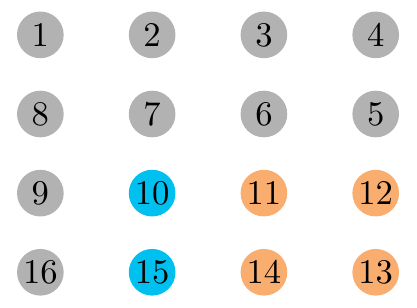}}
}
\caption{Parity basis of the columns. (a) The circuit maps the computational basis to the parity basis. (b) The parities stored in system qubit 10 (blue) and the corresponding ancilla (orange).
}
\end{figure}

\begin{figure}[htb]\label{fig:parity_basis_b}
\subfloat[]{\label{fig:cz_s2_b}
\includegraphics[height=3.2cm]{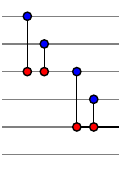}}\hspace{1cm}
\subfloat[]{\label{fig:cz_s2_c}
\includegraphics[height=3.2cm]{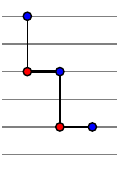}}
\caption{Equivalent circuits to Fig.~\ref{fig:cz_s2_a}. \protect\subref{fig:cz_s2_b} The ancilla qubits are in the parity basis and the system qubits are in the original basis. \protect\subref{fig:cz_s2_c} Both the system and the ancilla qubits
are in the parity basis.}
\end{figure}

The problem with implementing the circuit in Fig.~\ref{fig:many_hop_a} is that there are many nonlocal gates in the first and second parts, which we replot in Fig.~\ref{fig:cz_s2_a}.
To deal with this difficulty, we go to the parity basis of columns by applying the circuit in Fig.~\subref*{fig:add_parity} to both the system and the ancilla qubits. 
In this new basis, any system qubit stores the total parity of the original system qubits in the same column from the current location to the bottom; e.g., the qubit 10 in Fig.~\subref*{fig:parity_example} stores the parity $\sys_{10}+\sys_{15}$ in the parity basis of the columns. Any ancilla qubit stores the total parity to the lower left of the corresponding system qubit; e.g., the ancilla qubit corresponding to system qubit 10 in Fig.~\subref*{fig:parity_example} stores the parity $\sys_{11}+\sys_{12}+\sys_{13}+\sys_{14}$. To find the circuit corresponding to Fig.~\ref{fig:cz_s2_a} in the new basis, we first go to the parity basis of the ancilla qubits and keep the system qubits unchanged. With this intermediate basis, the first four (last two) pairs of \CZ\ gates in Fig.~\ref{fig:cz_s2_a} are mapped to the first (second) pair of \CZ\ gates in Fig.~\subref*{fig:cz_s2_b}. 
This is the case because we can combine all \CZ\ gates acting on the same system qubit into a single \CZ\ gate by using an ancilla qubit storing the total parity of the original ancilla qubits. We continue to go to the parity basis of the system qubits, and the circuit in Fig.~\subref*{fig:cz_s2_b} is mapped to the one in Fig.~\subref*{fig:cz_s2_c}. This is the case because the total parity of the  first and second (third and fourth) qubits in Fig.~\subref*{fig:cz_s2_b} is equal to the total parity of the first and third (third and fifth) qubits. 

In general, we have a \CZ\ gate between the system qubit on each odd-numbered row and the ancilla qubit two rows below it and a \CZ\ gate between the system and the ancilla qubits on each odd-numbered row except for the first row. Going to the parity basis of the system qubits might seem to be unnecessary, but we will show that it is essential for reducing the circuit depth. 

We have shown that the first two parts in Fig.~\ref{fig:many_hop_a} can be implemented with only local \CZ\ gates; however, going to the parity basis requires a sequence of $O(\sqrt \nso\ssp\ssp)$ gates which increases the circuit depth. One solution is to implement, in parallel, the quantum circuit in Fig.~\subref*{fig:add_parity} to all columns of system qubits at the beginning. As the ancilla qubits move to the left, they pick up the parities stored in the system qubits. These add up to the desired parities of the ancilla qubits in the parity basis, and no basis transformation is needed when the ancilla qubits move across the system qubits. The circuit in Fig.~\subref*{fig:cz_s2_c} is implemented in each time step. After the ancilla qubits reach the left end, we go back to the original basis by applying the gates in Fig.~\subref*{fig:add_parity}, in reverse order, to both the system and the ancilla qubits. We then move the ancilla qubits to the right by reversing the order of the \CNOT\ and the \SWAP\ gates when we move them to the left. The \CZ\ gates in the last part of Fig.~\ref{fig:many_hop_a} are implemented in each time step. All of the ancilla qubits are disentangled with the system qubits when they reach the right end, and the unitary $\Gamma$ has been implemented on the system qubits. We verify our procedure numerically by using random classical bit strings $\sysbf$ and $\sysbf'$ for system sizes up to $100\times 100$ and by simulating the actual quantum circuit for the system size $4\times 4$.

In summary, the unitary $\Gamma$ can be implemented with the following four stages:
\begin{enumerate}
 \item Implement the circuit in Fig.~\subref*{fig:add_parity} for each column of the system qubits.
 \item Move the ancilla qubits to the left while applying the \CZ\ gates in Fig.~\subref*{fig:cz_s2_c}.
 \item Undo the circuit in Fig.~\subref*{fig:add_parity} for both the system qubits and the ancilla qubits.
 \item Move the ancilla qubits to the right while applying the \CZ\ gates in the third part of Fig.~\ref{fig:many_hop_a}.
\end{enumerate}
Each of the four stages can be implemented with $O(\nso\ssp\ssp)$ gates and $O(\sqrt\nso\ssp\ssp)$ depth; therefore, the whole procedure takes gates and depth with the same scalings. By comparison, $\mathcal F_x$ or $\bF{y}$ requires $O(\nso^{1.5}\ssp)$ gates and $O(\sqrt\nso\ssp\ssp)$ depth to implement. Our quantum algorithm provides an example in which the parity problem in simulating 2D fermionic systems can be circumvented with negligible overhead.

\section{The Fermi-Hubbard model}
\label{sec:fhm}

One way to unravel the intricate physics in strongly correlated materials is to approximate them with idealized model Hamiltonians, such as the Fermi-Hubbard model. The Hubbard model captures many signatures of the physical systems, although it is too simple to describe real materials quantitatively. It has resisted a full solution despite decades of intense analytical and numerical studies. The Hubbard model has a relatively small number of interacting terms, allowing for easy implementation on a quantum computer. The single-band Hubbard model is described by the Hamiltonian
\begin{align}\label{eq:fhh_2}
\mathcal H_\mathrm{FH} = &- \sum_{\langle j,k \rangle, \sigma} t_{jk}\big( \c^\dagger_{j,\sigma} \c^{}_{k,\sigma} +\hc\big) +U \sum_j \num_{j,\uparrow}\num_{j,\downarrow}\nonumber\\
&\;\;\; + \sum_{j,\sigma} (\epsilon_j-\mu) \num_{j,\sigma}- \sum_j h_j (\num_{j,\uparrow}-\num_{j,\downarrow})\,,
\end{align}
where the first term represents fermions hopping between sites, the second term represents on-site interactions, the third term is a local potential field, and the last term represents a local magnetic field. 
 
\begin{figure}[htb]\label{fig:phase_diagram}
\centering
\includegraphics[width=0.45\textwidth]{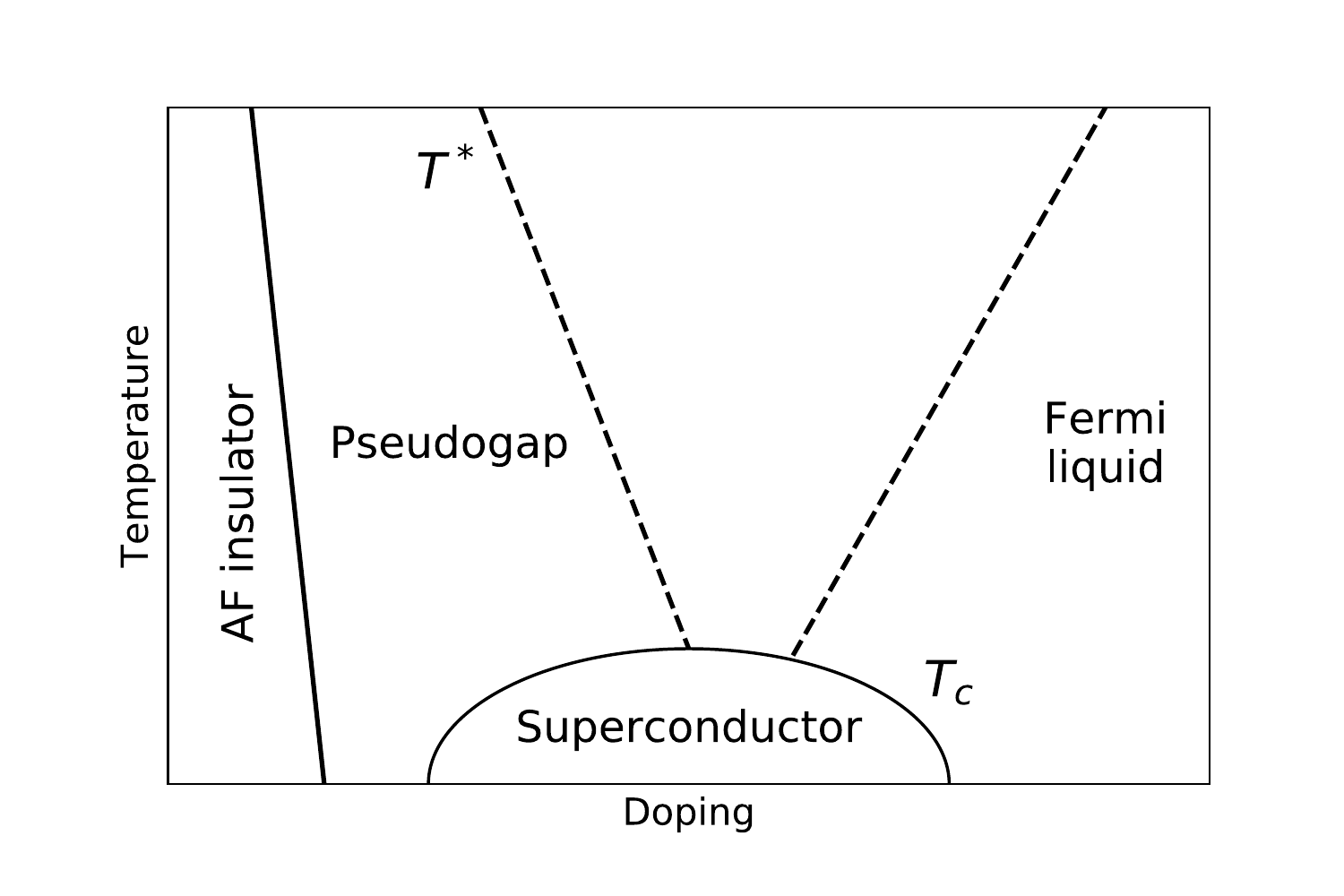}
\caption{Phase diagram of the 2D Fermi-Hubbard model. The critical temperature $T_c$ depends on the value of hole doping, AF stands for antiferromagnet.}
\end{figure}

The 2D Hubbard model is widely used as a canonical microscopic model for strongly correlated fermionic systems. It is believed to be a key ingredient in understanding the mechanism behind high-temperature superconductivity~\cite{dagotto_correlated_1994}. In particular, understanding the physics of the model in the intermediate interaction strength regime $U/t\sim 4 $ in the underdoped region remains an open problem. Multiple orders exist in this region of its phase diagram; see Fig.~\ref{fig:phase_diagram}. 
Even the nature of the ground state remains ambiguous due to the competition between a number of different order parameters~\cite{zheng_stripe_2016}. Significant progress has been made in the identification of the unconventional Mott-insulator transition and a superconducting phase with $d$-wave order parameter~\cite{halboth_renormalization-group_2000,maier_systematic_2005}. Recently, the antiferromagnet phase of the Fermi-Hubbard model was realized in optical lattices of about 80 sites at a temperature of 1/4 times the tunneling energy~\cite{mazurenko_cold-atom_2017}. This region of multiple competing phases is also the most interesting regime from the point of view of modeling materials; therefore, a simulation using even a relatively small quantum computer could provide alternative qualitative and quantitative insights into the physics of the model~\cite{Lloyd1997,Lloyd1999,wecker_solving_2015}. Generalizations of the Hubbard model beyond the single-band case and including more complicated lattices could be a route to modeling a wide range of strongly correlated materials. It is also a versatile tool for exploring strongly correlated electron phases~\cite{SachdevLandscape} in a controlled way. Simulation of the Hubbard model on a quantum computer could allow quantitative analysis of the physical characteristics beyond the phase diagram, such as dynamical effects which could help uncover the deeper physics of the strongly correlated phases in the model. 

\begin{figure}[htb]\label{fig:sublattice}
\centering
\includegraphics[width=0.32\textwidth]{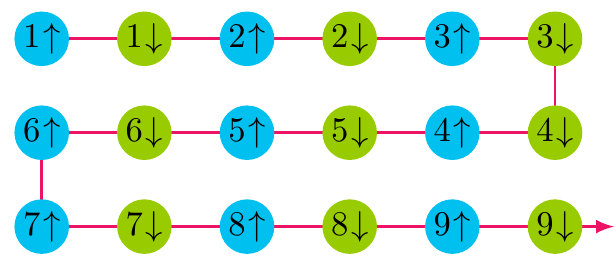}
\caption{One way to map the two spin states to qubits on a 2D qubit array using the JWT.}
\end{figure}

In order to implement time evolution of the Hubbard model on a quantum computer, we decompose it into a product of available quantum gates based on the Trotter-Suzuki formula~\cite{trotter_product_1959,suzuki_generalized_1976}; the desired accuracy determines the number of time steps~\cite{babbush_chemical_2015,poulin_trotter_2015}. We simulate the Trotter error numerically for small system sizes in Appendix~\ref{sec:n_trotter}. In each time step, we implement successively the horizontal hopping terms, the vertical hopping terms, and the remaining terms. The two spin states can be mapped to two sublattices of a 2D qubit array; for an example, see Fig.~\ref{fig:sublattice}. 
The horizontal hopping terms can be implemented by bringing qubits corresponding to the same spin states next to each other with the fermionic \SWAP\ gates; for example, one can swap the qubits $j\!\uparrow$ with $j\!\downarrow$ for an odd $j$ value in a row to implement some hopping terms, swap back, and then swap $j\!\uparrow$ with $j\!\downarrow$ for an even $j$ value in the same row to implement the remaining hopping terms. This step requires $O(\nso)$ gates and $O(1)$ depth. The vertical hopping terms can be implemented using the method described in Sec.~\ref{sec:fourier}; this requires $O(\nso)$ gates and $O(\sqrt \nso)$ depth. Alternatively, the vertical hopping terms can be implemented using the method described in Appendix~\ref{sec:move_ancilla}, also with $O(\nso)$ gates and $O(\sqrt \nso)$ depth. The on-site interaction terms in Eq.~\eqref{eq:fhh_2} are mapped to qubit operators of the form
\begin{align}
n_{j \uparrow} n_{j \downarrow} \,\mapsto\, \frac{1}{4}\left(I-Z_{j \uparrow}\right)\left(I-Z_{j \downarrow}\right)\,,
\end{align}
which can be implemented using a controlled-phase gate, 
\begin{gather}\label{eq:c_phase}
\exp\Big(\mathord-i \tau U n_{j \uparrow} n_{j \downarrow}\Big) \,\mapsto\, \left(
\begin{array}{cccc}
 1 & 0  & 0  & 0  \\
0  & 1 & 0  & 0  \\
0  & 0  & 1 & 0  \\
0  & 0  & 0  & e^{-i \tau U} \\
\end{array}
\right).
\end{gather}
The bias term and the magnetic term in the second line of Eq.~\eqref{eq:fhh_2} can be implemented straightforwardly with single-qubit operators. This step requires $O(\nso)$ gates and $O(1)$ depth. Putting all of these steps together, each Trotter step can be simulated with only $O(\nso)$ quantum gates and $O(\sqrt \nso)$ depth. 

Cooper proved that an arbitrarily small attraction between electrons can cause a pairing of electrons, leading to a lower energy state than the Fermi energy. As described by the BCS theory~\cite{bardeen_microscopic_1957}, the $s$-wave pairing wave function due to electron-phonon interactions is responsible for conventional superconductivity. The cuprate superconductivity has $d$-wave symmetry \cite{wollman_experimental_1993, tsuei_pairing_2000}, wherein the superconducting wave function changes sign upon rotation by 90\textdegree. It has been predicted that $d$-wave pairing underlies high-$T_c$ superconductivity in cuprates~\cite{scalapino_case_1995}; however, the mechanism of this pairing is not completely known. The mean-field Hamiltonian that describes $d$-wave pairing in the single-band Hubbard model is
\begin{align}
\hspace{-2pt}\mathcal{H}_\mathrm{DW} =& -\!\sum_{\langle j,k \rangle, \sigma} t_{jk} \left(\c^\dag_{j,\sigma}\c^{}_{k,\sigma} + \c^\dag_{k,\sigma}\c^{}_{j,\sigma} \right) - \mu \sum_{j, \sigma}\num_{j, \sigma}\nonumber\\
&\;\, -\sum_{\langle j,k \rangle} \Delta^{{x^2} - y^2}_{jk}\left(\c^\dag_{j\uparrow} \c^\dagger_{k\downarrow} - \c^\dag_{j\downarrow} \c^\dagger_{k\uparrow}  + \hc\right),
\label{eq:dwave_mf}
\end{align}
where the chemical potential term with $\mu$ regulating the total number of particles and $\Delta^{x^2- y^2}_{jk}=\pm \Delta/2$ are the superconducting gaps for the horizontal and vertical directions, respectively. With the mapping in Fig.~\ref{fig:sublattice}, the pairing term $\c^\dag_{j\uparrow} \c^\dagger_{k\downarrow}  + \hc$ can be implemented similarly to the hopping terms.

Assuming translational symmetry and periodic boundary conditions, the ground state of the Hamiltonian~\eqref{eq:dwave_mf} can be prepared using the fermionic Fourier transformation,
\begin{align}
 \c^\dag_{\kbf,\sigma} = \frac{1}{\sqrt{\nn_x\nn_y}} \, \sum_{j=1}^{\nn_x \nn_y} e^{2\pi i( k_x x_j+k_y y_j)} \c_{j,\sigma}^\dagger\,,
\end{align}
where $\nn_{x}$ $(\nn_{y})$ is the number of sites in the horizontal (vertical) direction, and $x_j\in \{1,\ldots,\nn_{x}\}$ and $y_j\in \{1,\ldots,\nn_{y}\}$ denote the coordinates of the $j$th site. The discrete wave vector $\kbf = (k_x,\, k_y)$ satisfies the condition $k_{x}\ssp\nn_{x} \in \{1,\ldots,\nn_{x}\}$, and similarly for $k_y$. In the momentum basis, the Hamiltonian~\eqref{eq:dwave_mf} becomes 
\begin{align}
\mathcal{H}_\mathrm{DW}=& \sum_{\kbf,\sigma} \xi_\mathbf{k}\, \c^\dag_{\kbf,\sigma} \c^{}_{\kbf,\sigma}  -\sum_{\kbf}  \Delta_\mathbf{k} \big(\c^\dag_{\kbf \uparrow} \c^{\dagger}_{-\kbf \downarrow} + \hc\big),
\label{eq:dwave_mf_q}
\end{align}
where $-\kbf \equiv (1-k_x, 1-k_y)$ and
\begin{gather}
 \xi_\mathbf{k} = -2t\, [\cos (2\pi k_x)+\cos (2\pi k_y)]-\mu\,,\\[4pt]
 \Delta_\mathbf{k} = \Delta \, [\cos (2\pi k_x)-\cos (2\pi k_y)]\,,
\end{gather}
where $\mu$ is chosen such that $\xi_\mathbf{k} = 0$ at the Fermi surface. The BCS mean-field ground state of Eq.~\eqref{eq:dwave_mf_q} is
\begin{gather}\label{eq:BCS_ground}
\ket{\Psi_\mathrm{DW}} =\prod_\mathbf{k}\left(u_\mathbf{k}\ssp  + v_\mathbf{k}\ssp \c_{\mathbf{k} \uparrow }^\dagger \c_{-\mathbf{k}\downarrow }^\dagger\right) \ket{\vac}\,,\\
u_\mathbf{k}^2=\frac{1}{2}\left(
1+\frac{\xi_\mathbf{k}}{\sqrt{\xi_\mathbf{k}^2+\norm{\Delta_\mathbf{k}}^2}}
\right)\,,\\[4pt]
v_\mathbf{k}^2=\frac{1}{2}\left(1-\frac{\xi_\mathbf{k}}{\sqrt{\xi_\mathbf{k}^2+\norm{\Delta_\mathbf{k}}^2}}\right)\,,
\end{gather}
where $u_\mathbf{k}\geq 0$ and $\sgn v_\mathbf{k} = \sgn \Delta_\mathbf{k}$. Inspired by Ref.~\cite{verstraete_quantum_2009}, we prepare the BCS ground state in the site basis by first preparing the ground state in the momentum basis~\eqref{eq:BCS_ground} and then applying the 2D Fourier transformation to the two spin states independently. 

\begin{figure}[htb]\label{fig:bogoliubov}
\centering
\includegraphics[width=0.37\textwidth]{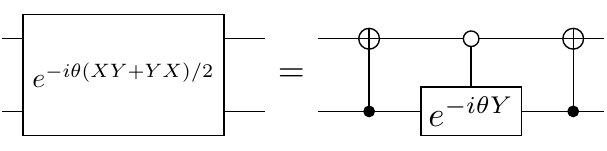}
\caption{Quantum circuit to implement the Bogoliubov transformation in the BCS state.}
\end{figure}

The ground state in the momentum basis can be prepared by applying a Bogoliubov transformation to the vacuum state:
\begin{align}\label{eq:bogo}
\ket{\Psi_\mathrm{DW}} = \prod_\mathbf{k} \exp\Big(\theta_\mathbf{k}\ssp\c^\dag_{\kbf\uparrow} \c^\dag_{-\kbf\downarrow}-\hc\Big)\, \ket{\vac}\,,
\end{align}
where $\sin \theta_\mathbf{k} = v_\mathbf{k}$. The corresponding generator is
\begin{align}
i\ssp\big(\c^\dag_{\kbf\uparrow} \c^\dag_{-\kbf\downarrow}-\hc\big) \,\mapsto\, \frac{1}{2}\, \Big(X_{\kbf\uparrow} Y_{-\kbf\downarrow} + Y_{\kbf\uparrow} X_{-\kbf\downarrow}\Big)\,,
\end{align}
where the qubit corresponding to the spin orbital $-\kbf\!\downarrow$ is put next to that of $\kbf\!\nsp\uparrow$ in the JWT. This unitary can be implemented with the quantum circuit in Fig.~\ref{fig:bogoliubov}.
The fermionic Fourier transformations, different for the two spin states due to the opposite orders in $\kbf$, can be performed using the method described in Sec.~\ref{sec:fourier}. The BCS mean-field state with periodic boundary conditions can thus be prepared using $O(\nso^{1.5})$ gates and $O(\sqrt\nso)$ circuit depth. 

An alternative way to prepare the $d$-wave mean-field state is by introducing the fermionic operators corresponding to the real single-particle wave functions,
\begin{gather}
 \c^\dag_{\kbf +,\sigma} = \frac{1}{\sqrt 2}\Big(\c^\dag_{\kbf,\sigma}+ \c^\dag_{-\kbf,\sigma}\Big)\,,\\[2pt]
 \c^\dag_{\kbf -,\sigma} = \frac{-i}{\sqrt 2}\Big(\c^\dag_{\kbf,\sigma}- \c^\dag_{-\kbf,\sigma}\Big)\,.
\end{gather}
for $k_1,k_2\leq 1/2$. With these operators, the Hamiltonian~\eqref{eq:dwave_mf_q} takes the form,
\begin{align}
\mathcal{H}_\mathrm{DW}&=\!\!\! \!\sum_{k_1,k_2 \leq 1/2,\,\sigma}\!\! \xi_\mathbf{k}\, \Big(\c^\dag_{\kbf+,\sigma} \c^{}_{\kbf+,\sigma} +\c^\dag_{\kbf-,\sigma} \c^{}_{\kbf-,\sigma}\Big) \nonumber\\
&\;\,-\!\!\!\!\sum_{k_1,k_2 \leq 1/2} \!\!\! \Delta_\mathbf{k} \Big(\c^\dag_{\kbf+ \uparrow} \c^{\dagger}_{\kbf+\downarrow} + \c^\dag_{\kbf- \uparrow} \c^{\dagger}_{\kbf-\downarrow} + \hc\Big)\,,
\label{eq:dwave_mf_r}
\end{align}
where the pairing terms are also ``diagonalized'' as a consequence of the real transformation matrix. Therefore, one can prepare the mean-field ground state by first preparing the Bogoliubov ground state of the spin orbitals $\kbf\pm\!\uparrow$ and $\kbf\pm\!\downarrow$ before performing the same real basis transformation for the two spin states. This method might be more efficient than the one using plane waves by avoiding the phase rotations.

In experiments, it is often the case that open boundary conditions are used. In this case, the horizontal or vertical hopping terms in Eq.~\eqref{eq:dwave_mf} correspond to a real triangular matrix with real eigenstates. We use these eigenstates as the single-particle basis states instead of the plane waves; therefore, the basis transformation matrix is also real. Similar to the case of periodic boundary conditions, the basis states on a 2D array with open boundary conditions can also be decomposed into a product of 1D basis states in each direction. This factorized form allows for efficient implementation of the 2D basis transformation with $O(\nso^{1.5})$ gates and $O(\sqrt \nso)$ depth using the method described in Sec.~\ref{sec:fourier}. When the Hamiltonian~\eqref{eq:dwave_mf} does not satisfy translational symmetry, we can still prepare the mean-field ground state using the method described in Sec.~\ref{sec:gaussian}. However, we might not be able to take advantage of the factorized form of the transformation, and it takes $O(\nso^{2})$ gates and $O(\nso)$ depth to prepare the mean-field ground state in the worst case.
 
The ground state of the 2D Fermi-Hubbard Hamiltonian~\eqref{eq:fhh} can be prepared by first preparing the mean-field ground state and then slowly interpolating from $H_\mathrm{DW}$ to  $H_\mathrm{FH}$~\cite{PRB2011FidelityNumerics}. Following Ref.~\cite{wecker_solving_2015}, we introduce the Hamiltonian
\begin{align}
\mathcal H(s)&=(1-s)\ssp \mathcal H_\mathrm{DW} +s\ssp\ssp\mathcal H_\mathrm{FH}\nonumber\\
&\;\, -\zeta\sum_{\langle j,k \rangle} \Delta^{{x^2} - y^2}_{jk}\left(\c^\dag_{j\uparrow} \c^\dagger_{k\downarrow} - \c^\dag_{j\downarrow} \c^\dagger_{k\uparrow}\right) + \hc
\nonumber\\
&\quad\; -\eta\!\sum_{\langle\langle j,k \rangle\rangle} i\Delta^{xy}_{jk}\left(\c^\dag_{j\uparrow} \c^\dagger_{k\downarrow} - \c^\dag_{j\downarrow} \c^\dagger_{k\uparrow}\right)  + \hc
\,,
\end{align} 
where $s$ slowly changes from $0$ to $1$ in the adiabatic algorithm and $\Delta_{jk}^{xy} = \Delta/2$ for $x_j-x_k = \pm 1$ and $y_j-y_k = \pm 1$. The coefficients $\zeta$ and $\eta$ introduce  small gaps to avoid quantum fluctuations of the $d$-wave order parameter and otherwise gapless nodal quasiparticles, respectively.

The initial Hamiltonian $H_\mathrm{DW}$ favors a mean-field ground state with $d$-wave symmetry of the pairing order parameter. The algorithm proceeds in the following way. We initialize the system in a specific mean-field wave function of the $d$-wave type,
\begin{gather}
\mathcal H_\mathrm{DW}\,\ketb{\Psi_\mathrm{DW}(s=0)} = E_0(s=0)\,\ketb{\Psi_\mathrm{DW}(s=0)},
\end{gather}
and adiabatically deform it to the ground state of the Hubbard model. If the initial wave function does not reflect the symmetry of the Fermi-Hubbard ground state, the adiabatic trajectory encounters a quantum phase transition. This transition manifests as a small spectral gap between the ground state and the excited states at the point of transition, which vanishes in the thermodynamic limit. If the initial mean-field state reflects the symmetry of the ground state of the Hubbard model, there is no phase transition and the spectral gap remains independent of the system size in the course of the evolution. One indication that the phase transition has occurred in the course of the evolution is that the final state contains excitations above the pairing gap, which can be detected by measuring the correlation functions. The controlled-phase gate~\eqref{eq:c_phase} is implemented off resonantly to achieve the required high fidelity in the Xmon superconducting qubits~\cite{barends_digitized_2016}. The optimal length of the gate translates into the limited strength of the effective interaction corresponding to roughly $U\sim 10\,\text{MHz}$ for the Xmons. For the desired parameter regime $U\sim 4t$, an upper bound on the superconducting gap can be inferred from typical values of the order parameters obtained numerically, $\Delta \lesssim 0.04t\sim 0.01U\sim 0.1\,\text{MHz}$.

\section{Summary}

In this paper, we discuss quantum simulation of strongly correlated fermionic systems using qubit arrays. We improve on an existing quantum algorithm to prepare an arbitrary Slater determinant~\cite{wecker_solving_2015,kivlichan_quantum_2017} by exploiting a unitary symmetry. We also present a quantum algorithm to prepare an arbitrary fermionic Gaussian state with $O(\nso^2)$ gates and $O(\nso)$ circuit depth. This algorithm---unlike existing ones that rely on translational symmetry---is completely general and is useful for simulating disordered systems and quantum impurity models. Our quantum algorithms are optimal in the sense that the number of parameters in the quantum circuit is equal to that describing the quantum states. We implement these algorithms as a part of the open-source project OpenFermion~\cite{mcclean_openfermion_2017}. 

We also present an algorithm to implement the 2D fermionic Fourier transformation on a 2D qubit array with $O(\nso^{1.5})$ gates and $O(\sqrt \nso)$ circuit depth, both of which scale better than methods based on fermionic \SWAP\ gates~\cite{babbush_low_2017}. A crucial step to achieve this optimal scaling is a unitary transformation that attaches the parity operators to the hopping terms; we show that it can be implemented with $O(\nso)$ gates and $O(\sqrt \nso)$ circuit depth. This approach provides an example where the parity problem in simulating fermionic systems in more than one spatial dimension can be circumvented with almost no additional cost. Our algorithm can also be used for any 2D transformation that factorizes into horizontal and vertical terms, such as the fermionic Fourier transformation with open boundary conditions. 

Using the Fermi-Hubbard model as an example, we discuss how to use our algorithms to find the ground-state properties and phase diagrams of strongly correlated systems. We demonstrate that the $d$-wave pairing mean-field states of the model can be prepared with $O(\nso^{1.5})$ gates and $O(\sqrt \nso)$ depth. We show that each Trotter step in the time evolution can be implemented with $O(\nso)$ gates and $O(\sqrt \nso)$ depth with only local qubit interactions. We discuss how to prepare the ground state of the model by adiabatically evolving the system from the mean-field Hamiltonian to the Hubbard Hamiltonian. The methods that we have develop can also be used for other quantum lattice models that suffer from the negative-sign problem, e.g., frustrated spin systems, the $t-J$ model, or lattice gauge theories. 

In conclusion, we show in this paper that physical properties of strongly correlated fermionic systems can be simulated on available 2D qubit arrays with only $O(\nso^{1.5})$ gates and $O(\sqrt \nso)$ circuit depth with very little overhead by using the Jordan-Wigner transformation. This result is one more step towards the goal of using quantum computers to investigate correlated quantum systems that are beyond the reach of any classical computer.

\section*{Acknowledgments}

The authors would like to acknowledge the enlightening and useful discussions with Ryan Babbush, Garnet Kin-Lic Chan, Chunjing Jia, Jarrod McClean, Andre Petukhov, Yaoyun Shi, Norman Tubman, and Fang Zhang. This work is supported by the NASA Advanced Exploration Systems program and the NASA Ames Research Center. The research is based in part upon work supported by the Office of the Director of National Intelligence (ODNI). K.J.S acknowledges support from NSF Grant No.~1717523. K.K. acknowledges support by NASA Academic Mission Services, contract number NNA16BD14C. All the quantum circuits in this work are plotted using the $\langle \textsf{q} \vert \textsf{pic} \rangle$ system developed by Thomas Draper and Samuel Kutin. The views and conclusions contained herein are those of the authors and should not be interpreted as necessarily representing the official policies or endorsements, either expressed or implied, of the ODNI or the U.S. Government. The U.S. Government is authorized to reproduce and distribute reprints for Governmental purposes notwithstanding any copyright annotation thereon. 

% \bibliographystyle{apsrev4-1_with_title}
% \bibliography{fhm,2DFHLit} 

%merlin.mbs apsrev4-1.bst 2010-07-25 4.21a (PWD, AO, DPC) hacked
%Control: key (0)
%Control: author (72) initials jnrlst
%Control: editor formatted (1) identically to author
%Control: production of article title (1) required
%Control: page (0) single
%Control: year (1) truncated
%Control: production of eprint (0) enabled
%

\appendix

\section{Quadratic Hamiltonians}
\label{sec:quadratic}

Hamiltonians that are quadratic in fermionic creation and annihilation operators are important in the mean-field descriptions of many-body quantum systems. The most general form of a quadratic Hamiltonian is
\begin{align}\label{eq:MF}
 \mathcal H = \sum_{j, k=1}^{\nso} \big(\mm_{jk}-\mu\ssp \delta_{jk}\big)\ssp \c_{j}^\dagger \c_{k} + \half\sum_{j,k=1}^{\nso} \Big(\Delta_{jk}\ssp \c_{j}^\dagger \c_{k}^\dagger +\hc\Big)\,,
\end{align}
where $\mm = \mm^\dagger$ and $\Delta = - \Delta^T$ are complex matrices and the chemical potential $\mu$ regulates the total number of particles; hereafter, $\mu$ will be absorbed into $\mm$ to simplify notation. We review the standard results on how to bring $\mathcal H$ into the diagonal form
\begin{align}\label{eq:dia_MF}
 \mathcal H = \sum_{j=1}^{\nso} \varepsilon_{j}\ssp\b_{j}^\dagger \b_{j} + \cnumber\,, 
\end{align}
where $\b_j$ and $\b_j^\dagger$ are a new set of fermionic operators that satisfy the canonical anticommutation relations.

When the number of particles is conserved ($\Delta = 0$), the Hamiltonian~\eqref{eq:MF} takes the form
\begin{align}\label{eq:quadratic_N_conv}
  \mathcal H = \sum_{j, k=1}^{\nso} \mm_{jk}\ssp \c_{j}^\dagger \c_{k}  \,.
\end{align}
The commutator of $\mathcal H$ and a single creation operator is
\begin{align}
 \commut{\mathcal H}{\c_l^\dagger} &= \sum_{jk} M_{jk} \commut{\c_j^\dagger \c_k}{\c_l^\dagger} \nonumber\\
 &= \sum_{jk} M_{jk}\ssp \c_j^\dagger \ssp\delta_{kl} = \sum_j M_{jl}\ssp \c_j^\dagger\,.\label{eq:commut_Hm}
\end{align}
% $\mathbf{\c}^T = (\c_{1}\cdots \c_{\nso})$ 
In matrix form, we have
\begin{align}
 \commut{\mathcal H}{\mathbf \c^\dagger} = \mm^T \mathbf \c^\dagger = \mm^* \mathbf \c^\dagger\,,
\end{align}
where $\mathbf{\c}^\dagger = (\c_{1}^\dagger\cdots \c_{\nso}^\dagger)^T$. The time evolution under the Hamiltonian $\mathcal H$ takes the form,
\begin{align}\label{eq:U_M}
 e^{-i \tau\mathcal H} \ssp\mathbf \c^\dagger e^{i \tau\mathcal H} = e^{-i\tau \mm^T}\nsp \mathbf \c^\dagger\,.
\end{align}
Therefore, the time evolution of the $2^{\nso}\times 2^{\nso}$ matrix $\mathcal H$ can be represented by the $\nso\times \nso$ unitary matrix $e^{-i\tau M^T}$. The quadratic form~\eqref{eq:quadratic_N_conv} can be diagonalized into the form~\eqref{eq:dia_MF} by introducing the fermionic operators $\b_j$ and $\b_j^\dagger$ such that
\begin{align}
 {\mathbf \b}^\dagger = U {\mathbf \c}^\dagger\,,\qquad U\nsp \mm^T U^\dagger = \diag(\varepsilon_1,\ldots, \varepsilon_{\nso})\,,
\end{align}
where $U$ is an $\nso\times\nso$ unitary matrix and the eigenvalues satisfy $\varepsilon_{1} \leq \varepsilon_{2}\cdots  \leq \varepsilon_{\nso}$. In the $\nel$-particle sector, the ground state of the Hamiltonian~\eqref{eq:quadratic_N_conv} corresponds to a Slater determinant with the $Q$ matrix being the first $\nel$ rows of $U$; see Eq.~\eqref{eq:slater}.  

When the number of particles is not conserved, we rewrite the Hamiltonian~\eqref{eq:MF} in matrix form,
\begin{align}\label{eq:MF_matrix}
 \mathcal H = \half
 \begin{pmatrix}
 \mathbf{\c}^\dagger& 
 \mathbf{\c} 
\end{pmatrix}
\begin{pmatrix}
 \Delta  & \mm\\[2pt]
 -\mm^* & -\Delta^*
\end{pmatrix}
\begin{pmatrix}
 \mathbf{\c}^\dagger\\[2pt]
 \mathbf{\c} 
\end{pmatrix}+\cnumber\,,
\end{align}
where the extra constant comes from different ordering of the fermionic operators; the matrix form~\eqref{eq:MF_matrix} is symmetrically ordered, while the form in Eq.~\eqref{eq:MF} is normally ordered. The standard way to diagonalize the Hamiltonian~\eqref{eq:MF_matrix} is by solving the Bogoliubov-de Gennes equations, which is somewhat cumbersome for numerical implementations. Alternatively, one introduces the Majorana fermion operators,
\begin{align}\label{eq:Majo_trans}
 \s_j =\sqhalf (\c_j^\dagger + \c_j)\,,\quad   \s_{j+\nso} = \frac{i}{\sqrt 2}( \c_j^\dagger - \c_j)\,,
\end{align}
for $j=1,2,\ldots,\nso$, which satisfy the anticommutation relations
\begin{align}\label{eq:Majo_commut}
 \anticommut{\s_j}{\s_k} = \delta_{j k},\,\;\; \text{for}\; j, k=1,2,\ldots,2\nso\,.
\end{align}
In matrix form, the transformation~\eqref{eq:Majo_trans} reads
\begin{align}
 \mathbf \s =  \Omega
\begin{pmatrix}
 \mathbf{\c}^\dagger\\[2pt]
 \mathbf{\c} 
\end{pmatrix}\,,\qquad 
\Omega = \frac{1}{\sqrt 2}
\begin{pmatrix}
        \mathbb 1 &  \mathbb 1\\[2pt]
        i\mathbb 1  &  -i\mathbb 1
     \end{pmatrix}\,,
\end{align}
where $\mathbf \s = (\begin{matrix}\s_1 & \s_2 & \cdots & \s_{2\nso}\end{matrix})^T$. The Hamiltonian~\eqref{eq:MF_matrix} can be rewritten in terms of the Majorana fermion operators:
\begin{align}\label{eq:MF_Majo}
 \mathcal H = \frac{i}{2}\, \mathbf \s^T\! A\, \mathbf \s+\cnumber\,,
\end{align}
where $A$ is a $2\nso\times 2\nso$ real antisymmetric matrix,
\begin{align}\label{eq:Majo_matrix}
 \qquad A = 
 -i\ssp\Omega^*
 \begin{pmatrix}
 \Delta  & \mm\\[2pt]
 -\mm^* & -\Delta^*
\end{pmatrix}
\Omega^\dagger\,.
\end{align}
Conversely, any real antisymmetric matrix $A$ corresponds to a quadratic Hamiltonian of the form~\eqref{eq:MF}. The commutator between a single Majorana fermion and the Hamiltonian $\mathcal H$ is
\begin{align}
 \commut{\mathcal H}{\s_l} 
 &= \frac{i}{2}\,\sum_{j\neq k} A_{jk}\,\commut{\s_j \s_k}{\s_l}
%  &= \frac{i}{2}\, \sum_{j\neq k} A_{jk} \big(\delta_{kn}\, \s_j - \delta_{jn}\, \s_k\big) 
 = i \sum_{j\neq l} A_{jl}\, \s_j\,.\label{eq:Majo_commut_MF}
\end{align}
Using the matrix form $\commut{\mathcal H}{\mathbf \s} = -i A \,\mathbf \s$, we have
\begin{align}
 e^{-i \tau\mathcal H} \ssp\mathbf \s\, e^{i \tau\mathcal H} = e^{-\tau A} \ssp\mathbf \s\,.
\end{align}
Therefore, the unitary evolution of the Hamiltonian $\mathcal H$ can be represented by the orthogonal matrix $e^{-\tau A}$. The matrix $A$ can be brought into the standard form (equivalent to the Schur form up to a permutation) by an orthogonal transformation $\orth$,
\begin{align}\label{eq:A_diag}
  \orth A \orth^T = 
 \begin{pmatrix}
 0  & \mathcal E\\[2pt]
 -\mathcal E & 0
\end{pmatrix}\,,\quad 
\mathcal E = \diag(\varepsilon_1,\ldots, \varepsilon_{\nso})\,,
\end{align}
where $0\leq \varepsilon_{1} \leq \varepsilon_{2}\cdots  \leq \varepsilon_{\nso}$. This transformation corresponds to a basis change of the Majorana fermions $ \mathbf \s' = \orth\ssp\ssp \mathbf \s$, where the new operators $\mathbf \s'$ also satisfy the anticommutation relations~\eqref{eq:Majo_commut}. Going back to the picture of creation and annihilation operators, such a basis transformation takes the form
\begin{align}\label{eq:operator_mapping}
\begin{pmatrix}
 \mathbf{\b}^\dagger\\
 \mathbf{\b} 
\end{pmatrix} 
 =  W
 \begin{pmatrix}
 \mathbf{\c}^\dagger\\
 \mathbf{\c} 
\end{pmatrix}\,,
\end{align}
where $(\mathbf{\b}^\dagger \; \mathbf{\b} )^T= (\b_{1}^\dagger\cdots\, \b_{\nso}^\dagger\; \b_{1}\cdots\,\b_{\nso})^T$ and $(\mathbf{\c}^\dagger \; \mathbf{\c} )^T= (\c_{1}^\dagger\cdots\, \c_{\nso}^\dagger\; \c_{1}\cdots\, \c_{\nso})^T$. The unitary matrix $W$ defines a new set of fermionic operators $\b_j$ and $\b_j^\dagger$, which brings the Hamiltonian~\eqref{eq:MF} to the diagonal form~\eqref{eq:dia_MF}. The matrix $W$ is related to $R$ through a basis transformation,
\begin{align}\label{eq:W_R}
W = \Omega^\dagger\! \orth\,\Omega \,.
\end{align}
It is the starting point of our procedure in Sec.~\ref{sec:gaussian}, which prepares the ground state of the Hamiltonian~\eqref{eq:MF} on a quantum computer. In summary, $W$ can be calculated in four steps:
\begin{enumerate}
 \item Write the quadratic Hamiltonian~\eqref{eq:MF} in the matrix form~\eqref{eq:MF_matrix}.
 \item Find the real antisymmetric matrix $A$ using the matrix form~\eqref{eq:MF_matrix} and Eq.~\eqref{eq:Majo_matrix}.
 \item Bring $A$ into the standard form~\eqref{eq:A_diag} using an orthogonal transformation $\orth$.
 \item Calculate the matrix $W$ using $\orth$ and Eq.~\eqref{eq:W_R}. 
\end{enumerate}

\section{Ancilla-assisted fermionic gates}
\label{sec:move_ancilla}

In Sec.~\ref{sec:fourier}, we discuss how to implement the 2D fermionic Fourier transformation with $O(\nso^{1.5})$ gates and $O(\sqrt\nso)$ circuit depth. A crucial ingredient of our method is a unitary that attaches the corresponding parity operators to the bare hopping terms. This procedure allows efficient implementation of both horizontal and vertical hopping terms on a 2D qubit array. The same strategy can be used to simulate Hamiltonian time evolution---again on a 2D qubit array---by Trotterization, where a single time step can be achieved with only $O(\nso)$ gates and $O(\sqrt \nso)$ depth. Here, we propose an alternative method to implement the hopping terms in a Trotter step using ancilla qubits to store the parities also with $O(\nso)$ gates and $O(\sqrt \nso)$ depth. However, it is less efficient in circuit depth to perform the 2D Fourier transformation compared to the method in Sec.~\ref{sec:fourier} due to difficulties in parallelization.

\begin{figure}[ht]\label{fig:2D_lattice}
\centering
% \subfloat[]{\label{fig:2D_lattice_a}
% \includegraphics[width=0.225\textwidth]{mancilla_e2.pdf}
% }\hspace{1pt}
% \subfloat[]{\label{fig:2D_lattice_b}
% \includegraphics[width=0.225\textwidth]{mancilla_e.pdf}
% } \\
\subfloat[]{\label{fig:2D_lattice_a}
\includegraphics[width=0.225\textwidth]{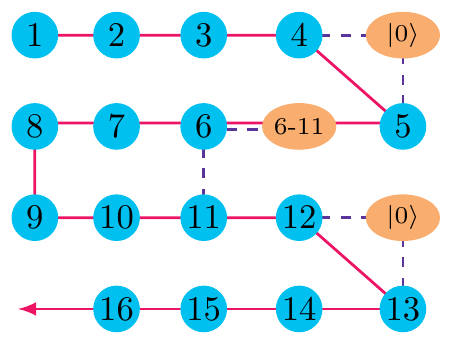}
}\hspace{1pt}
\subfloat[]{\label{fig:2D_lattice_b}
\includegraphics[width=0.225\textwidth]{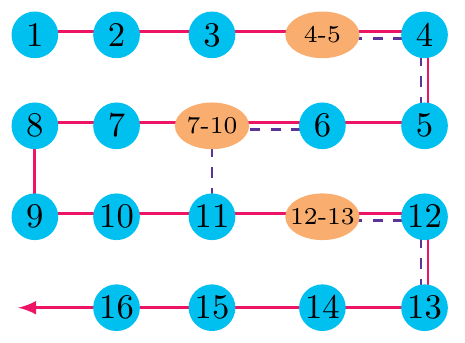}
}\\
\subfloat[]{\label{fig:2D_lattice_c}
\includegraphics[width=0.225\textwidth]{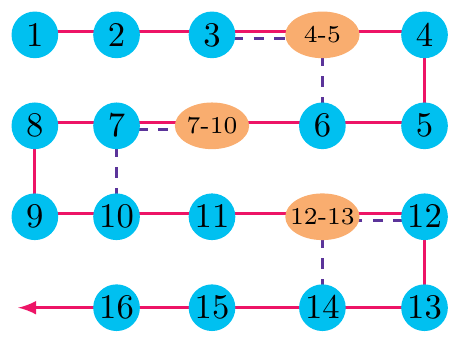}
}\hspace{1pt}
\subfloat[]{\label{fig:2D_lattice_d}
\includegraphics[width=0.225\textwidth]{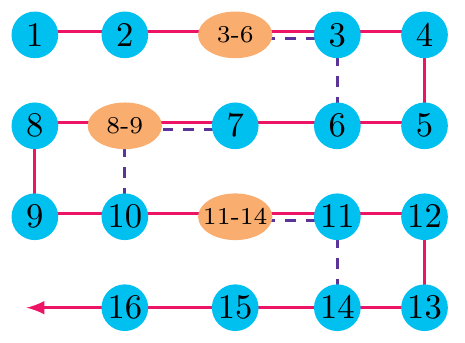}
}
\caption{\protect\subref{fig:2D_lattice_a}-\protect\subref{fig:2D_lattice_d} Implementing the hopping terms on a $4\times 4$ qubit array. The red arrow indicates the direction of the JWT and the system qubits (the blue circles) are numbered by their order in the JWT. The ancilla qubits (the orange ellipses) store the parity information of the system qubits, where $j\myhyphen k$ denotes the parity $Z_{j\myhyphen k}\equiv Z_j\cdots Z_k$. The purple dashed lines represent quantum gates between neighboring qubits.}
\end{figure}

\begin{figure}[htb]\label{fig:v_hop}
\centering
\includegraphics[width=0.45\textwidth]{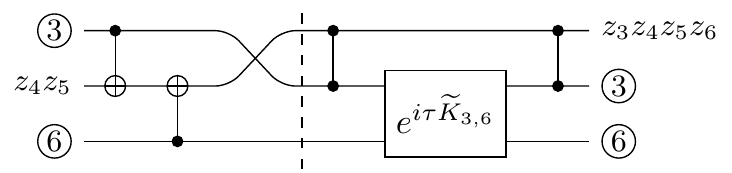}
\caption{Quantum circuit to implement the vertical hopping term between qubits 3 and 6.}
\end{figure}

We discuss how to implement the hopping terms on a 2D qubit array using the same example of array size $4\times 4$. The ancilla qubits move horizontally in the array to facilitate vertical hopping terms; see Fig.~\ref{fig:2D_lattice}.
The ancilla qubits are initially at the far left side of the qubit array; the states of those ancilla qubits for the right-closed hopping terms are set to $\ket{0}$, while those for the left-closed hopping terms store the parities of the two corresponding rows, e.g., the ancilla on the second row initially stores the parity $Z_{5\myhyphen 12}$. The vertical hopping terms are implemented with the parity stored in the ancilla qubits, whose states are updated constantly. For example, the vertical hopping term between qubits 3 and 6 in Fig.~\subref*{fig:2D_lattice_d} is
\begin{align}
% K_{3,6} = 
 \widetilde K_{3,6}\ssp Z_{4\myhyphen 5} =  -\widetilde K_{3,6}\ssp Z_{3\myhyphen 6}\,,\label{eq:vertical_hopping}
\end{align}
where $\widetilde K_{3,6} = (X_3 X_6 + Y_3 Y_6)/2$ denotes the bare hopping term. The time evolution $\exp(-i\tau \widetilde K_{3, 6})$ can be implemented using the second part of the circuit in Fig.~\ref{fig:v_hop}, where the parity stored in the ancilla qubit is attached to the bare hopping term $\widetilde K_{3,6}$ by the \CZ\ gates; 
the first part of the circuit, corresponding to Fig.~\subref*{fig:2D_lattice_c}, updates the parity and swaps the ancilla qubit with system qubit 3. The order of the two parts in Fig.~\ref{fig:v_hop} may be reversed for other hopping terms, e.g., the one between system qubits 7 and 10; see Figs.~\subref*{fig:2D_lattice_c} and \subref*{fig:2D_lattice_d}. After the ancilla qubits have swept over the entire rows, they are pushed back in the reverse direction and the same procedure starts over again. The vertical hopping terms require $O(\nso)$ two-qubit gates to implement for a single Trotter step. The horizontal hopping terms can be implemented straightforwardly as long as the two involved qubits are at the same side of the ancilla qubit (no need to update the parity).

In Appendix~\ref{sec:qubit_ladder}, we discuss how to simulate the 2D Fermi-Hubbard model with qubits on a ladder, i.e., two coupled chains. The basic idea is to use the fermionic \SWAP\ gate to change between row-major order and column-major order so that both the horizontal and vertical couplings can be implemented with only local interactions. This idea was also proposed by Kivlichan \textit{et al}.~\cite{kivlichan_quantum_2017} independently. Because such a strategy does not take full advantage of the 2D qubit interactions, the number of gates scales as $O(\nso^{1.5})$, which is worse than $O(\nso)$, by using the methods described above and in Sec.~\ref{sec:fourier}.

\section{Hamiltonian-based fermionic gates}
\label{sec:hamiltonian_based}

Using methods in quantum optimal control~\cite{werschnik_quantum_2007}, one can prepare the desired quantum states by applying sequences of Hamiltonian evolutions to initial states that are easy to prepare. When the fermionic Hamiltonians are quadratic in creation and annihilation operators, one can use their matrix representation to find the optimal control sequences. 

As a simple example, we discuss how to prepare the ground state of a hopping Hamiltonian on three orbitals, 
\begin{align}\label{eq:hopping_3}
\mathcal H_{J_{\nsp X}} = \frac{1}{\sqrt 2} \Big(\c^\dagger_{2} \c_{1} + \c^\dagger_{3} \c_{2} +\hc\Big)\,.
\end{align}
Besides $\mathcal H_{J_{\nsp X}}$, we also need $\mathcal H_{J_{\nsp Z}}$ in our control set,
\begin{align}
\mathcal H_{J_{\nsp Z}} = \c^\dagger_{1} \c_{1} - \c^\dagger_{3} \c_{3}\,,
\end{align}
whose ground states are easy to prepare. Using the normal ordered form in Eq.~\eqref{eq:quadratic_N_conv}, we have the matrix representations for $\mathcal H_{J_{\nsp X}}$ and $\mathcal H_{J_{\nsp Z}}$, 
\begin{align}
J_{\nsp X} =
\frac{1}{\sqrt 2} 
\begin{pmatrix}
 0& 1& 0\\[-2pt]
 1& 0& 1\\[-2pt]
 0& 1& 0
\end{pmatrix}\,,\qquad
J_{\nsp Z}  = 
\begin{pmatrix}
 1& 0& 0\\[-2pt]
 0& 0& 0\\[-2pt]
 0& 0& -1
\end{pmatrix}\,,
\end{align}
which are the same as the corresponding spin-1 angular momentum operators. Using two rotations along the $X$ axis and the $Z$ axis, we have
\begin{align}
 J_{\nsp X} = e^{-i\pi J_{\nsp Z}/2}e^{-i\pi J_{\nsp X}/2}\,J_{\nsp Z}\, e^{i \pi J_{\nsp X}/2}e^{i\pi J_{\nsp Z}/2} \,.
\end{align}
Using the isomorphism~\eqref{eq:U_M}, one can prepare eigenstates of the hopping Hamiltonian $\mathcal H_{J_{\nsp X}}$ by applying $\mathcal U$ to the corresponding eigenstates of $\mathcal H_{J_{\nsp Z}}$, where
\begin{align}
\mathcal U =  \exp\big(\mathord-i \pi\mathcal H_{J_{\nsp Z}}/2\big)\ssp \exp\big(\mathord-i \pi\mathcal H_{J_{\nsp X}}/2\big) \,.
\end{align}
Since $J_{\nsp Z}$ is diagonal, the eigenstates of $\mathcal H_{J_{\nsp Z}}$ are easy-to-prepare Slater determinants in the computational basis~\eqref{eq:hatree_fock}.

As a second example, we discuss the permutation of orbitals,
\begin{align}
 \begin{pmatrix}
 1& 2& 3\\[-2pt]
 3& 2& 1
\end{pmatrix}\,,
\end{align}
which requires three fermionic \SWAP\ (\FSWAP) gates on neighboring qubits in the JWT. We show how it can be achieved with a single multiqubit gate using Hamiltonian evolution by $\mathcal H_{J_{\nsp X}}$. The matrix powers of $J_{\nsp X}$ are
\begin{align}\label{eq:power_Jx}
J_{\nsp X}^2 = \half
\begin{pmatrix}
 1& 0& 1\\[-2pt]
 0& 2& 0\\[-2pt]
 1& 0& 1
\end{pmatrix}\,,\qquad 
J_{\nsp X}^3 = J_{\nsp X}\,.
\end{align}
From Eq.~\eqref{eq:power_Jx}, we can derive the exponential of $J_{\nsp X}$,
\begin{align}
 e^{-i\tau J_{\nsp X}} 
 %&= 1 + J_{\nsp X}\Big[-i\tau-\frac{(i\tau)^3}{3!}- \frac{(i\tau)^5}{5!}\cdots\Big]  + J_{\nsp X}^2 \Big[\frac{(i\tau)^2}{2!}+ \frac{(i\tau)^4}{4!}\cdots\Big]\nonumber\\[3pt]
 &= 1 - i\sin\tau J_{\nsp X} + (\cos\tau -1) J_{\nsp X}^2\,.
\end{align}
For $\tau=\pi$, we have
\begin{align}\label{eq:exp_Jx}
 e^{-i\pi J_{\nsp X}} = 1 - 2J_{\nsp X}^2 = -
 \begin{pmatrix}
 0& 0& 1\\[-2pt]
 0& 1& 0\\[-2pt]
 1& 0& 0
\end{pmatrix}\,.
\end{align}
Therefore, the unitary that swaps the first orbital with the third orbital is equal to
\begin{align}\label{eq:f_swap_3}
 \mathcal F_{\ssp\mSWAP}^{1,3} = \exp\bigg(\mathord-i\pi\sum_{j=1}^3\c^\dagger_{j} \c_{j} \bigg) \exp\big(\mathord-i\pi \mathcal H_{J_{\nsp X}}\big)\,,
\end{align}
where the first term on the right-hand side fixes the extra minus sign in Eq.~\eqref{eq:exp_Jx}. The two constituent parts in Eq.~\eqref{eq:f_swap_3} commute with each other, and the first part can be implemented with only single-qubit gates with the JWT.

\section{Fermionic SWAP gate}
\label{sec:FSWAP}

The \FSWAP\ gate on two neighboring qubits takes the form~\cite{verstraete_quantum_2009}
\begin{align}
 F_\mSWAP^{q,q+1} = \begin{pmatrix}
            1 &0 &0 &0\\
            0 &0 &1 &0\\
            0 &1 &0 &0\\
            0 &0 &0 &-1
           \end{pmatrix}\,,
\end{align}
where we use the standard order of basis states $\ket{00},\,\ket{01},\,\ket{10},\,\ket{11}$. One way to implement the fermionic SWAP gate is to use the \ISWAP\ gate 
\begin{gather}\label{eq:iswap}
  \mathrm{ISWAP}  =
\begin{pmatrix}
  1 &0 &0 &0\\
  0 &0 &i &0\\
  0 &i &0 &0\\
  0 &0 &0 &1
\end{pmatrix}= e^{(i\pi/2) H_\mISWAP^{q,q+1}}\,,\\
H_\mISWAP^{q,q+1} = \frac{1}{2}\, \big(X_q X_{q+1}+Y_q Y_{q+1}\big) \,.
\end{gather}
To fix the phases in \ISWAP, we introduce
\begin{align}
 e^{(i\pi/4)(Z_q + Z_{q+1})} = 
 \begin{pmatrix}
  i &0 &0 &0\\
  0 &1 &0 &0\\
  0 &0 &1 &0\\
  0 &0 &0 &-i
\end{pmatrix}\,,
\end{align}
which commutes with the \ISWAP\ gate. The fermionic \SWAP\ gate then takes the form
\begin{align}
 F_\mSWAP^{q,q+1} = -i\ssp e^{(i\pi/4)(Z_q + Z_{q+1})}\,e^{(i\pi/2) H_\mISWAP^{q,q+1}} \,.
\end{align}

\section{Simulating the 2D Hubbard model with a qubit ladder }
\label{sec:qubit_ladder}

\begin{figure}[htb]\label{fig:spin_ladder}
\centering
\includegraphics[width=0.47\textwidth]{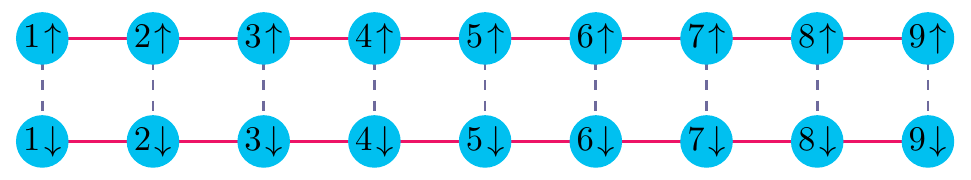}
\caption{A $2\times 9$ qubit ladder (two chains) is used to simulate the 2D Hubbard model on a $3\times 3$ lattice. The spin-up and -down orbitals are mapped to the upper and lower qubit chains, respectively, using two independent JWTs. The solid red lines represent nearest-neighbor interactions between qubits in the same JWT, and the dashed blue lines represent $ZZ$ interactions between qubits representing different spin states.}
\end{figure}

Experimentally, it is often easier to build qubits on a ladder (two coupled chains) than on a 2D lattice. The two spin chains in a ladder can be used to map fermionic modes with two different spin states; see Fig~\ref{fig:spin_ladder}. Because there are no hopping terms between the two spin states in the Hubbard model, the quantum state can be mapped using two independent JWTs on the two spin chains. One needs only to implement hopping terms within the chains and interaction terms between corresponding qubits in the two chains. In Ref.~\cite{reiner_emulating_2016}, the authors discussed physical implementations of analog quantum simulators of the 1D Hubbard model on a qubit ladder. Here, we discuss a digital quantum simulation of the 2D Hubbard model using the same architecture. The challenge is to implement both the horizontal and vertical hopping terms with only local qubit operations. We solve this problem by reordering the spin orbitals encoded in the JWT using the fermionic \SWAP\ gate (see Appendix~\ref{sec:FSWAP}). Similar ideas were proposed independently by Kivlichan \textit{et al.}~\cite{kivlichan_quantum_2017}.

\begin{figure}[htb]\label{fig:1D_chain}
\centering
\includegraphics[width=0.45\textwidth]{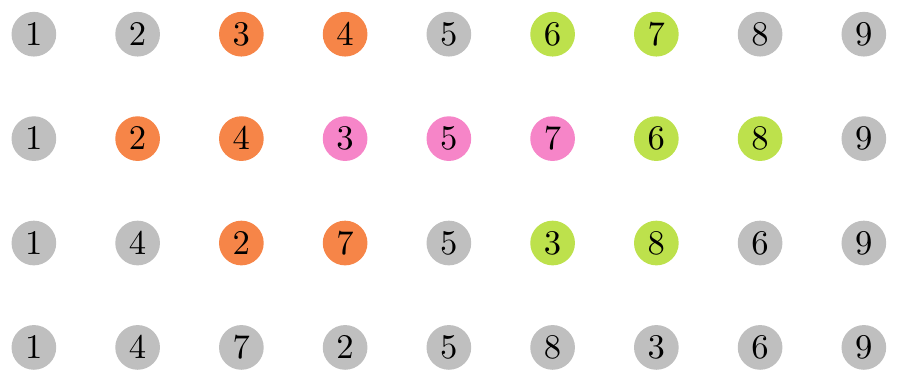}
\caption{The in situ matrix transposition~\eqref{eq:in_situ_transp} can be decomposed into elementary fermionic permutations involving only neighboring qubits (time goes from top to bottom). The first line is the initial ordering and the last line is the target ordering. In each line, the qubits plotted with the same bright color are involved in one elementary permutation, and their positions are updated in the next line. The gray qubits remain in the same place. The pink permutation can be implemented with three \FSWAP\ gates involving only neighboring qubits, or by Hamiltonian evolution, as described in Appendix~\ref{sec:hamiltonian_based}. The rest of the permutations require a single \FSWAP\ gate.}
\end{figure}

We demonstrate our approach using an example, in which the Fermi-Hubbard model on a $3\times 3$ lattice is mapped to a qubit ladder of size $2\times 9$; see Fig.~\ref{fig:spin_ladder}. The spin-up (-down) orbitals are mapped to the upper (lower) chain in the ladder using the JWT. 
To implement both the horizontal and vertical hopping terms, one can switch between the row-major and column-major orders,
\begin{align}
\begin{pmatrix}
 1& 2& 3\\
 4& 5& 6\\
 7& 8& 9
\end{pmatrix}
\xrightarrow{\text{transposition}}
\begin{pmatrix}
 1& 4& 7\\
 2& 5& 8\\
 3& 6& 9
\end{pmatrix}\,,
\end{align}
which corresponds to the following in situ transposition of the fermionic orbitals in the JWT:
\begin{align}\label{eq:in_situ_transp}
\begin{pmatrix}
 1& 2& 3& 4& 5& 6& 7& 8& 9\\
 1& 4& 7& 2& 5& 8& 3& 6& 9
\end{pmatrix}\,.
\end{align}
This permutation can be decomposed into elementary permutations involving only neighboring qubits; see Fig.~\ref{fig:1D_chain}. 
The pink permutation in Fig.~\ref{fig:1D_chain} can be implemented using three \FSWAP\ gates involving only neighboring qubits; in Appendix~\ref{sec:hamiltonian_based}, we discuss an alternative approach to implementing this permutation using Hamiltonian evolution. The number of two-qubit gates required to implement the in situ transposition for the two chains is $N_\mathrm{trans} = 2\times 9=18$. The numbers of two-qubit gates needed for the on-site interaction terms and the hopping terms are $N_\mathrm{int} = 9$ and $N_\mathrm{hop} = 2\times 2\times 6 = 24$, respectively. The total number of two-qubit gates needed to simulate a single Trotter step is $N_\mathrm{Trott} = N_\mathrm{trans} + N_\mathrm{int} + N_\mathrm{hop} = 51$. 

To end this appendix, we consider simulating the 2D Fermi-Hubbard model on a lattice of dimension $\nn_x\times \nn_y$ using a qubit ladder of length $\nn_x\ssp \nn_y$. Without loss of generality, we assume that $\nn_y \geq \nn_x$. The fermionic orbitals are mapped to qubits in row-major order using the JWT. Implementing the in situ matrix transposition on the entire chain requires $O(\nso^2)$ two-qubit gates and is inefficient. Instead, we can implement the in situ matrix transposition on orbitals in neighboring rows. This modification allows one to implement all of the hopping terms on a lattice with only $O(\nso^{1.5})$ two-qubit gates. For example, a single Trotter step can be implemented with seven constituent operations:
\begin{enumerate}
 \item Implement the horizontal hopping terms and the on-site interaction terms.
 \item Implement in situ transpositions on spin orbitals on the first and second rows, third and fourth rows, and so on.
 \item Apply vertical hopping terms between these pairs of rows.
 \item Undo the in situ transposition.
 % \item apply the horizontal hopping terms;
 \item Implement in situ transposition on spin orbitals on the second and third rows, fourth and fifth rows, and so on.
 \item Apply vertical hopping terms between these pairs of rows.
 \item Undo the in situ transposition.
\end{enumerate}
A single in situ transposition involving two rows requires $(\nn_x-1)\nn_x/2$ \FSWAP\ gates on neighboring qubits to implement. Because there are $4(\nn_y-1)$ in situ transpositions in one Trotter step, the total number of two-qubit gates required is $N_\mathrm{trans} = 2(\nn_y-1)(\nn_x-1)\nn_x = \nso \nn_x-\nso -2\nn_x^2+2\nn_x$, where $\nso = 2\nn_y\nn_x$ is the total number of spin orbitals. The numbers of two-qubit gates needed for the on-site interaction terms and the hopping terms are $N_\mathrm{inter} = \nso/2$ and $N_\mathrm{hop} = 2(\nn_y-1)\nn_x+2(\nn_x-1)\nn_y = 2\nso -2 (\nn_y+\nn_x)$, respectively. For square lattices, the total number of two-qubit gates needed for a single Trotter step is $N_\mathrm{Trott} = N_\mathrm{trans} + N_\mathrm{inter} + N_\mathrm{hop} \simeq  \nso^{3/2}/\sqrt 2+\nso/2$. For example, the total number of two-qubit gates is $275$ for $50$ spin orbitals on a square lattice with a size of $5\times 5$. 

\section{Numerical study of the Trotter error}
\label{sec:n_trotter}

One way to prepare strongly correlated quantum states is by adiabatic evolution. On a digital quantum computer, we can simulate the evolution by dividing it into discrete time steps and approximating the Hamiltonian within each time step. Here, we numerically study the errors that arise from using a second-order Trotter formula to simulate adiabatic state preparation for the ground states of the Fermi-Hubbard model.

For simplicity, we consider only the hopping terms:
\begin{align}
  \Hhop = - \sum_{\langle j,k \rangle, \sigma} t_{jk}
  \big( c^\dagger_{j,\sigma} c_{k,\sigma} +\hc\big)\,, 
\end{align}
and the interaction term
\begin{align}
  \Hint = U \sum_j n_{j,\uparrow}n_{j,\downarrow} -
  \mu \sum_{j,\sigma} \nn_{j, \sigma}\,.
\end{align}
We introduce a chemical potential term in $\Hint$ to fix the particle number of the ground state so that it occurs at half filling. Since the ground state of $\Hhop$ is also half filled and both $\Hhop$ and $\Hint$ conserve particle number, the particle number stays constant throughout the evolution. To prepare a ground state of $\Hint$ starting from a ground state of $\Hhop$, we apply the time-dependent Hamiltonian
\begin{align}
 \mathcal H(s) = (1 - s) \Hhop + s\ssp \Hint,
  \label{tdham}
\end{align}
where $s=t/T \in [0,\,1]$, with $T$ being the total evolution time. If we split $T$ into $n$ time steps, then a single Trotter step takes the form
\begin{align}
  e^{-i (1-s)\ssp \Delta t\ssp \Hhop} e^{-i s\ssp \Delta t\ssp \Hint},
  \label{trotter_step}
\end{align}
where $\Delta t = T/n$.

\begin{figure}[htb]
\label{fig:fidelity_a}
\includegraphics[width=0.9\linewidth]{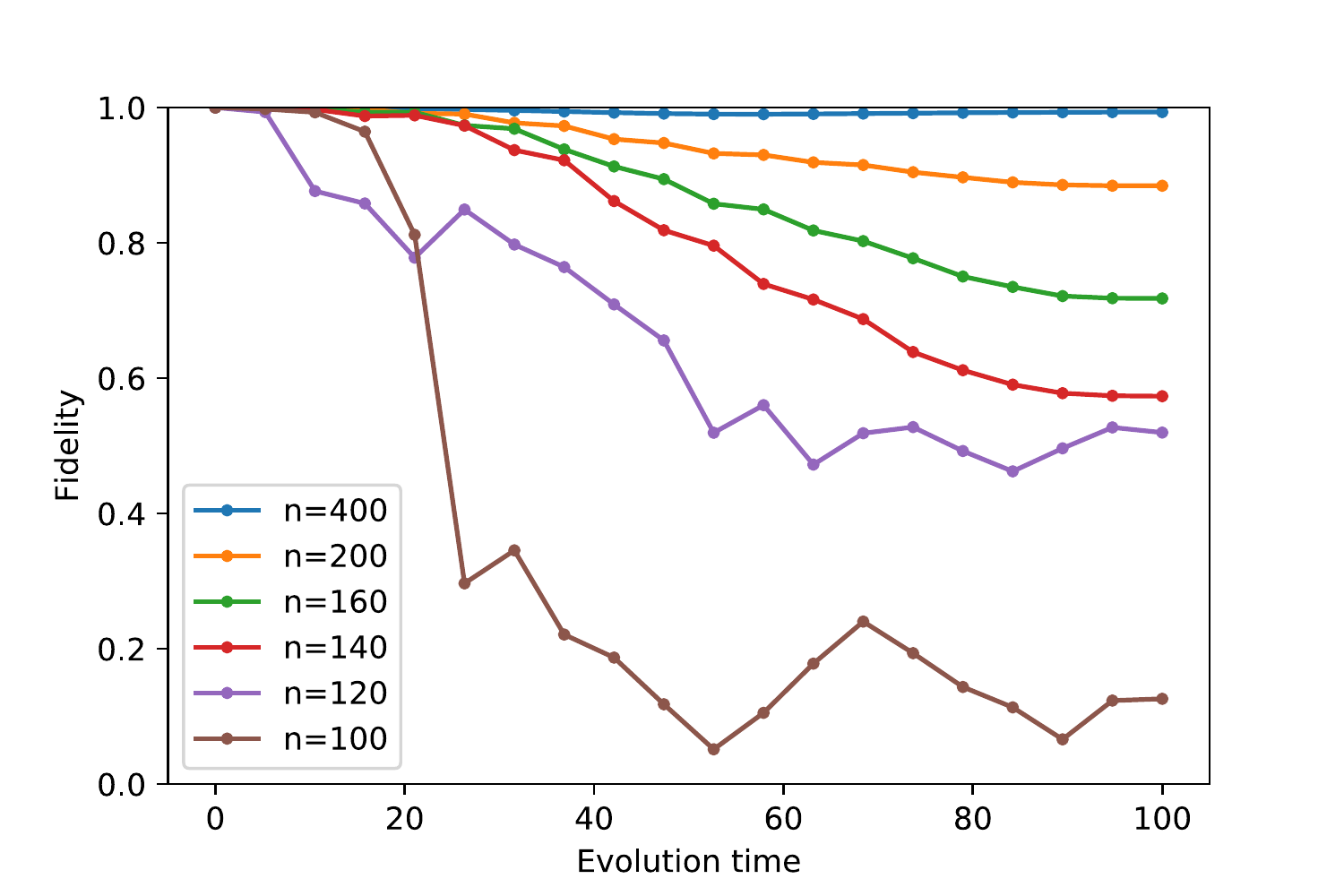}
\caption{Fidelity with states from adiabatic evolution for
  various Trotter step numbers on a $2\times 2$ grid.}
\end{figure}

\begin{figure}[htb]\label{fig:fidelity_b}
\includegraphics[width=0.9\linewidth]{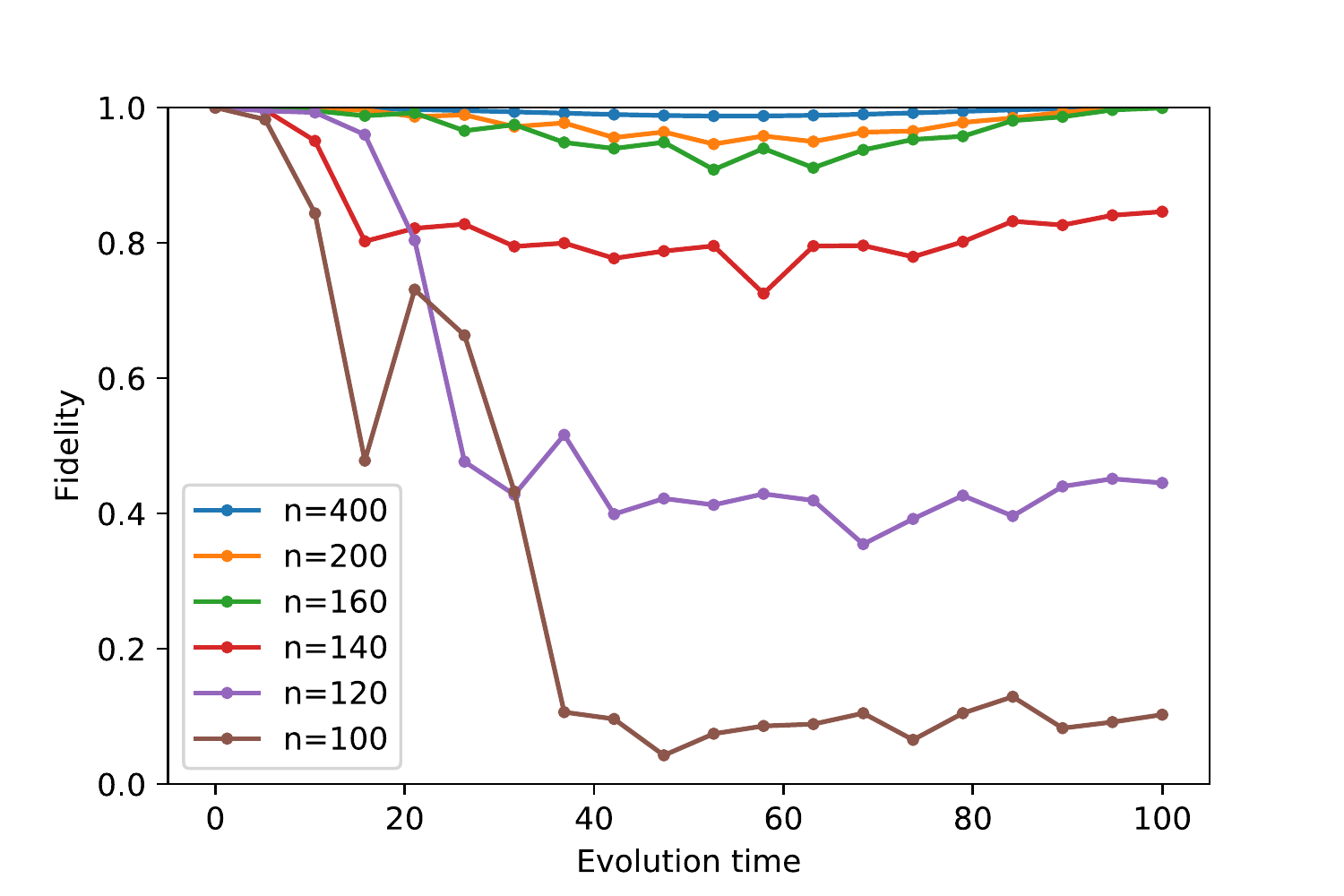}
\caption{Fidelity with states from adiabatic evolution for
  various Trotter step numbers on a $3\times 2$ grid.}
\end{figure}

\begin{figure}[htb]\label{fig:fidelity_c}
\includegraphics[width=0.9\linewidth]{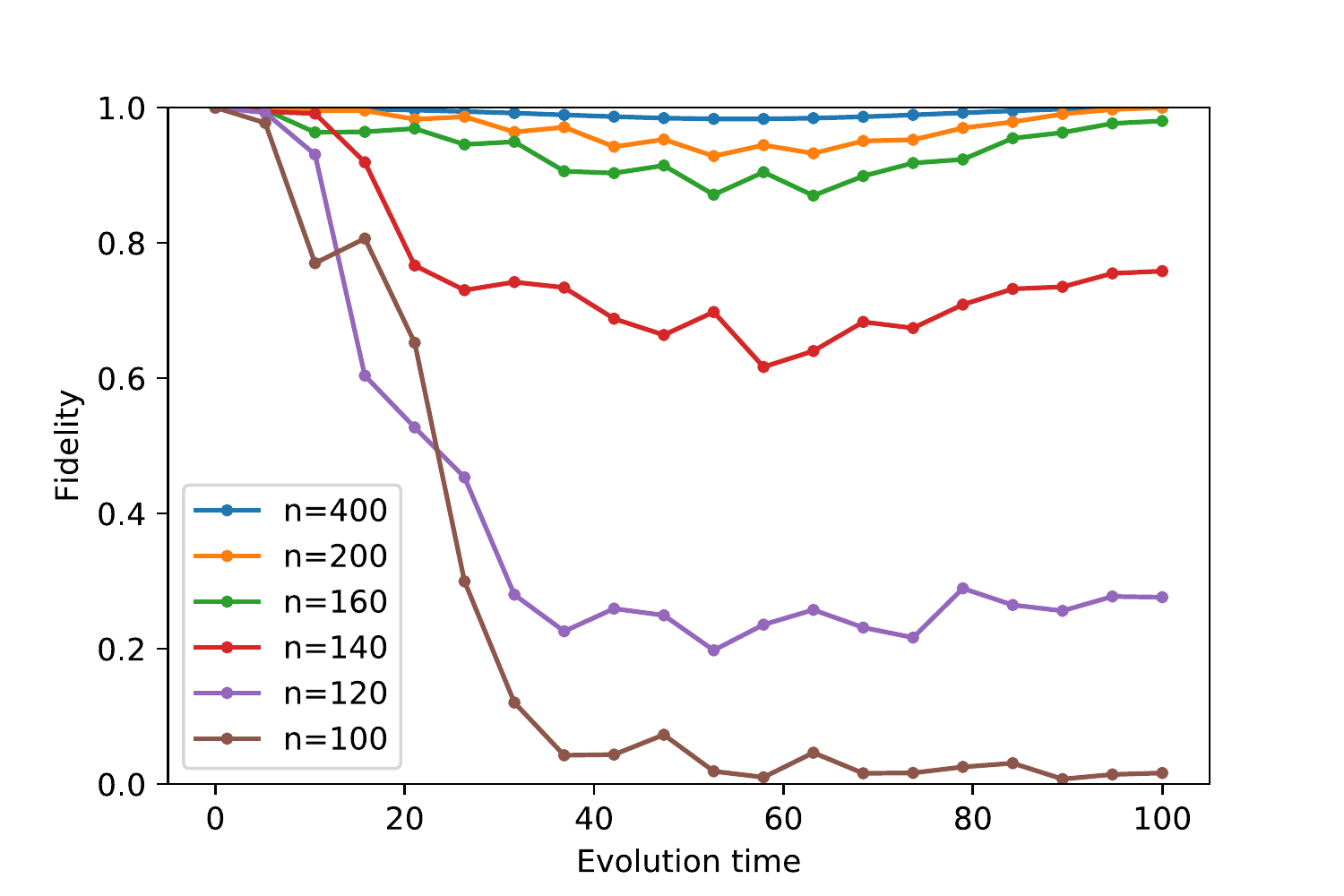}
\caption{Fidelity with states from adiabatic evolution for
  various Trotter step numbers on a $4\times 2$ grid.}
\end{figure}

For the purposes of this paper, we ignore the issue of approximating the two time-evolution operators in Eq.~(\ref{trotter_step}). We use the parameter settings
\begin{gather}
  t_{jk} = 1 \hspace{5pt} \text{for all cases of $\langle j,k \rangle$},\quad   U = 4, \quad \text{and}\;\,   \mu = 1,
  \label{params}
\end{gather}
on two-dimensional grids of various sizes. We use the open-source package OpenFermion~\cite{mcclean_openfermion_2017} to initialize the Hamiltonians of the Hubbard model and compute their ground states. Then, we use the open-source package QuTiP~\cite{johansson_qutip_2013} to compute the exact evolution by the time-dependent Hamiltonian (\ref{tdham}), starting with the ground state of $\Hhop$. We adjust the total evolution time $T$ until the final state is within the ground space of $\Hint$, indicating that the evolution is adiabatic. For all of the grid sizes in our work, $T = 100$ is sufficient. Finally, we approximate the evolution using the second-order Trotter formula with various values of $n$, the number of steps. Higher values of $n$ give a better approximation. For each step size, we record the quantum state 20 times throughout the evolution and compute the fidelity of these states with the states from the true adiabatic evolution. The numerical results for grid sizes $2\times 2$, $3\times 2$, and $4\times 2$ are shown in Figs.~\ref{fig:fidelity_a}, \ref{fig:fidelity_b}, and \ref{fig:fidelity_c}, respectively. 

\section{Rearranged quantum circuits for different hopping terms}
\label{sec:circuits_Gamma} 

\begin{figure}[htb]\label{fig:many_hop_a2}
\includegraphics[height=3.2cm]{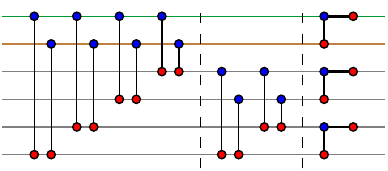}
\caption{The \CZ\ gates between the system (blue) and ancilla (red) qubits in a single time step.}
\end{figure}
\begin{figure}[htb]\label{fig:many_hop_b2}
\includegraphics[height=3.2cm]{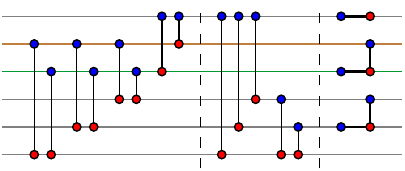}
\caption{Rearranged circuit for hopping terms between the second and third rows.}
\end{figure}
\begin{figure}[htb]\label{fig:many_hop_c}
\includegraphics[height=3.2cm]{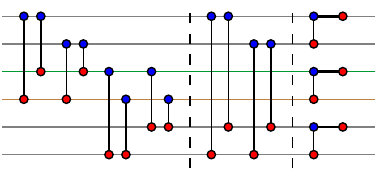}
\caption{Rearranged circuit for hopping terms between the third and fourth rows.}
\end{figure}
\begin{figure}[htb]\label{fig:many_hop_d}
\includegraphics[height=3.2cm]{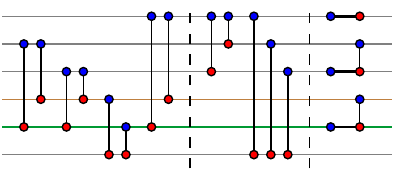}
\caption{Rearranged circuit for hopping terms between the fourth and fifth rows.}
\end{figure}
\begin{figure}[htbp]\label{fig:many_hop_e}
\includegraphics[height=3.2cm]{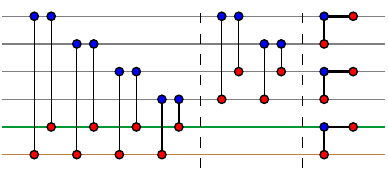}
\caption{Rearranged circuit for hopping terms between the fifth and sixth rows.}
\end{figure}
In Sec.~\ref{sec:fourier}, we provide an algorithm to implement the 2D fermionic Fourier transformation with $O(\nso^{1.5})$ gates and $O(\sqrt\nso\ssp\ssp)$ depth. A crucial step is a unitary transformation $\Gamma$, diagonalized in the computation basis, that introduces the desired parities to the bare vertical hopping terms. Therein, we provide a method to implement $\Gamma$ with $O(\nso\ssp\ssp)$ gates and $O(\sqrt\nso\ssp\ssp)$ depth by introducing one ancilla per row to store the parities of the system qubits. In each time step, the ancilla qubits interact with system qubits in a particular column with the circuit in Fig.~\ref{fig:many_hop_a2}. The first and second parts of the circuit consist of \CZ\ gates between each system qubit and the ancilla qubits below, starting with the next odd-numbered row. This circuit is used to demonstrate the hopping terms between the first and second rows, and a rearranged circuit in Fig.~\ref{fig:many_hop_b2} is used for hopping terms between the second and third rows. Here, we provide the other arrangements of the same circuit for the remaining pairs of rows that are adjacent to each other. The circuit in Fig.~\ref{fig:many_hop_a2} can be transformed into those in Figs.~\ref{fig:many_hop_c} and~\ref{fig:many_hop_e} by regrouping neighboring pairs of gates, which can be generalized to any right-closed hopping term for a system with an even number of rows. Similarly, the first three pairs of gates and the last two gates in the second part in Fig.~\ref{fig:many_hop_b2} can be regrouped into gates in the corresponding positions in Fig.~\ref{fig:many_hop_d}, which can also be generalized to any left-closed hopping term for larger system sizes. Each of the remaining gates in Fig.~\ref{fig:many_hop_b2} (except for the third part) involve the first system qubit and an ancilla qubit below; these gates do not affect any left-closed hopping term, either.

\end{document}